\documentclass[12pt]{article}

\usepackage{amsmath,amssymb,color,graphics,amscd,amsfonts,mathrsfs,epsf,indentfirst}

\usepackage{bm}
\usepackage{subfigure}

\usepackage{multirow,graphicx}

\usepackage[numbers,sort&compress]{natbib}

\title{Dynamical Meson Melting in Holography}

\author{Takaaki Ishii$^a$, Shunichiro Kinoshita$^b$, Keiju Murata$^c$ and Norihiro Tanahashi$^{d,e}$\\ 
{\small \it $^{a}$Crete Center for Theoretical Physics, Department of Physics}\\
{\small \it University of Crete, PO Box 2208, 71003 Heraklion, Greece}\\
{\small \it $^b$ Osaka City University Advanced Mathematical Institute, Osaka 558-8585, Japan}\\
{\small \it $^c$ Keio University, 4-1-1 Hiyoshi, Yokohama 223-8521, Japan}\\
{\small \it $^d$ Kavli Institute for the Physics and Mathematics of the Universe (WPI)}\\
{\small \it The University of Tokyo, Kashiwa, Chiba 277-8583, Japan}\\
{\small \it $^e$ DAMTP, University of Cambridge, Wilberforce Road, Cambridge CB3 0WA, UK}\\
}

\begin{document}
  
\maketitle

\begin{abstract}
We discuss mesons in thermalizing gluon backgrounds in the
$\mathcal{N}=2$ supersymmetric QCD using the gravity dual. We
numerically compute the dynamics of a probe D7-brane in the Vaidya-AdS
geometry that corresponds to a D3-brane background thermalizing from
zero to finite temperatures by energy injection. In static backgrounds, 
it has been known that there are two kinds of brane embeddings where 
the brane intersects the black hole or not. They correspond to the phases 
with melted or stable mesons. In our dynamical setup, we
obtain three cases depending on final temperatures and injection
time scales. The brane stays outside of the black hole horizon when the final
temperature is low, while it intersects the horizon and settles down to the static
equilibrium state when the final temperature is high. Between these two
cases, we find the {\it overeager case} where the brane dynamically
intersects the horizon although the final temperature is not high enough
for a static brane to intersect the horizon. The interpretation of this
phenomenon in the dual field theory is meson melting due to
non-thermal effects caused by rapid energy injection.
In addition, we comment on the late time evolution of the brane and a possibility of its reconnection.
\\
\flushright{
AP-GR-108\\
CCTP-2014-26\\
CCQCN-2014-17\\
IPMU14-0004\\
OCU-PHYS-396
}
\end{abstract}

\newpage
\tableofcontents

\section{Introduction and Summary}
\label{sec:introandsum}

\subsection{Introduction}
\label{subsec:introduction}

The gauge/gravity duality~\cite{Maldacena:1997re,Gubser:1998bc,Witten:1998qj} enables us to tackle strongly-coupled gauge theories by using classical gravity. The deconfined phase in the finite-temperature gauge theories is described by the gravity dual in the presence of black holes~\cite{Witten:1998zw}. This viewpoint has been applied to studying characteristics of the quark-gluon plasma, observed at the RHIC and the LHC experiments. It has been predicted that in the strongly coupled plasma, the shear viscosity over entropy is universally very small~\cite{Policastro:2001yc,Son:2002sd,Policastro:2002se,Kovtun:2003wp,Buchel:2003tz,Kovtun:2004de}. The gravity computations are related with fluid dynamics~\cite{Baier:2007ix,Bhattacharyya:2008jc,Hubeny:2011hd}.

An advantage of using the gravity dual is that time-dependent systems can be easily handled. Thermalization and its out-of-equilibrium dynamics are of interest in real-world processes in experiments. In the gravity dual, thermalization is described by black hole formation~\cite{Bhattacharyya:2009uu}.\footnote{A generalization of~\cite{Bhattacharyya:2009uu} to inhomogeneous shell collapse is studied in~\cite{Balasubramanian:2013rva,Balasubramanian:2013oga}.
See also Ref.~\cite{Craps:2013iaa} for a generalization to a confining background.} 
The timescale for the thermalization is typically fast, and the plasma evolution can be understood from hydrodynamic computations~\cite{Janik:2006gp,Benincasa:2007tp,Beuf:2009cx,Chesler:2009cy,Heller:2011ju,Heller:2012je}. On the other hand, very-far-from-equilibrium effects can be significant in the very beginning of the fast thermalization. We would like to consider such a situation in QCD using the gravity dual.

In this paper, we will study time-dependent dynamics of mesons in fast thermalization in the gauge/gravity duality. In particular, we would like to focus on the initial-stage effects in the thermalization. For that purpose, we will do numerical computations to solve the partial differential equations describing time evolution in the gravity dual. As the setup, we will consider the $\mathcal{N}=2$ SQCD realized by D-branes.

The $\mathcal{N}=2$ SQCD provides a simple ground for catching qualitative features of QCD-like theories. This theory is realized by $N_c$ D3-branes and $N_f$ D7-branes, where the D7-branes supply fundamental quarks. In the strong coupling in the large-$N_c$ limit, the D3-branes are replaced with the AdS geometry, and we probe the curved background with D7-branes in the limit of $N_f \ll N_c$ \cite{Karch:2002sh}. In this paper, we consider the case when $N_f=1$.%
\footnote{
We may alternatively consider the case that $N_f>1$ and the $N_f$ D7-branes oscillate coherently.
}
In finite temperatures, the gravity background is the AdS-Schwarzschild black hole with flat horizon~\cite{Witten:1998zw}. There is no confinement/deconfinement phase transition in the color sector of the $\mathcal{N}=2$ SQCD, and the gluons are deconfined in finite temperatures. In the flavor sector, however, if we introduce a nonzero quark mass to explicitly break the global $U(1)$ symmetry of the theory, we can find two phases with or without stable mesons, where the parameter is the ratio of the quark mass and the temperature.

The phases are characterized by the embeddings of the D7-branes into the black hole geometry~\cite{Babington:2003vm}. There are two kinds of embeddings when the system is static. Let us consider units where the quark mass is unity; the parameter is then the temperature determined by the black hole size.
In low temperatures, the D7-branes do not touch the black hole horizon, and the fluctuations on the D7-branes give stable mesons~\cite{Kruczenski:2003be}. This case is called the Minkowski embedding. In high temperatures, the D7-branes end on the black hole, and the quasi-normal modes on the D7-branes corresponds to melting mesons in the thermal background~\cite{Hoyos:2006gb,Myers:2007we}. This case is called the black hole embedding. The phase transition has been studied in detail in~\cite{Kirsch:2004km,Ghoroku:2005tf,Apreda:2005yz,Mateos:2006nu,Albash:2006ew,Mateos:2007vn}.\footnote{%
This phase transition was firstly noticed in the chiral-symmetry breaking D$4$/D$6$ system in~\cite{Kruczenski:2003uq}.}

Our interest is what happens when we consider dynamical embeddings of the D7-brane in dynamical backgrounds. 
To address this question, we will compute time evolutions of the D3/D7 system by 
numerically solving the equations of motion, which are 
given by partial differential equations.\footnote{
See Ref.~\cite{Evans:2010xs} for an earlier study on non-linear dynamics in the D3/D7 system.
}
 We consider dynamical embeddings of the probe D7-brane in a D3-brane background corresponding to a system thermalizing from zero to finite temperatures.

\subsection{Setup}
\label{subsec:setup}

Let us introduce our setup in a more concrete manner.
In this paper, we take the Vaidya-AdS$_5\times S^5$ spacetime as the gravity background.
In the gravity dual, thermalization is realized by the gravitational collapse and the subsequent black hole formation. The Vaidya-AdS metric is one of the simplest geometry of this kind, obtained for instance as a result of planar-shell collapse~\cite{Bhattacharyya:2009uu} or the collapse of null dust.%
\footnote{The Vaidya-AdS metric has been used for studying thermalization in holography, where non-local operators such as the entanglement entropy, spatial geodesic, and Wilson loop are considered~\cite{Hubeny:2007xt,AbajoArrastia:2010yt,Albash:2010mv,Balasubramanian:2010ce,Balasubramanian:2011ur}.}
This collapse in the bulk gravity will be identified with a spatially-homogeneous injection of a fixed amount of energy per unit volume in the boundary field theory.
We focus on the simplest case of the Vaidya-AdS metric to clarify basic aspects of the dynamical behaviors of the holographic mesons, though
in general we could consider more complicated thermalization patterns such as those realized by local heating caused by point particle infalling into the bulk black hole~\cite{Amsel:2007cw} or an anisotropic wake induced by a time-dependent deformation of the boundary metric~\cite{Chesler:2008hg}.
The thermalization time scale is typically fast for the Vaidya-AdS metric~\cite{Garfinkle:2011hm,Garfinkle:2011tc,Wu:2012rib},
and hence it would be suitable to focus on the effect of rapid thermalization.

\begin{figure}[tbp]
\begin{center}
\includegraphics[width=4cm]{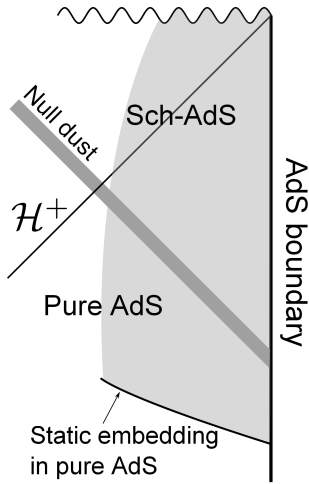}
\end{center}
\caption{The Penrose diagram of the Vaidya-AdS$_5$ spacetime.
The initial data for the bulk brane configuration is given by the static embedding in the pure
AdS. The null dust is injected to generate the bulk black hole, and $\mathcal{H}^+$ is the event horizon of the black hole. The world-volume of the brane is shown by the light gray region. In this figure, the final state of the D7-brane embedding is the black hole embedding.
}
\label{ponchi}
\end{figure}

The time evolution we consider is sketched as follows. In the
Vaidya-AdS$_5\times S^5$ spacetime, it is convenient to use the ingoing
Eddington-Finkelstein time $V$ as the bulk time coordinate, and $V$ coincides
with the time of the field theory at the AdS boundary. Initially for $V<0$, the spacetime is locally pure AdS. We then inject energy from the AdS boundary at $V=0$ to form the black hole that settles down to the Schwarzschild-AdS$_5$ black hole in the final state.
The initial static embedding of the D7-brane before the energy injection is given by the analytic solution in the pure AdS spacetime.
After the injection, the brane starts to change its embedding shape. We will numerically compute the evolution of the branes. Figure~\ref{ponchi} gives schematic illustration when the final mass of the black hole is sufficiently large and the brane intersects with the black hole horizon.

Here, we would like to emphasize subtleties in treating dynamical backgrounds of this kind. When we discuss non-stationary phenomena in the gravity dual description, especially when the event horizon exists, we should pay careful attention to the causality and non-uniqueness of the time foliation of the bulk spacetime, unlikely to static or stationary cases.
In our study, such subtleties associated with the bulk causality become manifest when we try to answer questions such as the following:
\begin{itemize}
\item How should we determine the phase of the brane embeddings at a given time? 
\item How can we know the moment at which the phase transition takes place in the boundary theory?
\end{itemize}
In static cases, the former question becomes almost trivial: 
We can distinguish the BH and Minkowski embeddings just by seeing whether the brane intersects with the event horizon or not.
In general dynamical cases, however, this naive consideration is no longer suitable. Firstly, there is no definitely preferred time foliation,
and this fact implies that the brane configuration at a given bulk time 
depends on bulk time foliation and is not unique.
In addition, the part of the brane world-volume captured in the event horizon cannot be observed by any observers outside the black hole, since that part lies in the causal future of those observers.
The information of that part does not affect the dynamics of the other part of the brane outside the black hole, as well as time-dependent boundary observables determined by the brane dynamics. 

This observation tells us that some naive notions based on the static cases are no longer rigorous in the dynamical cases.
We address this issue in more detail in Sec.~\ref{sec:results}.
In the gauge/gravity duality, it is the dual boundary operators that contains information of the boundary field theory. In our study, we will compute the time evolution of the operators by solving the equation of the motion of the brane. This approach guarantees that the results are not affected by bulk time foliations.

\subsection{Summary}
\label{subsec:summary}

Below, we summarize properties of the dynamical D7-brane embeddings clarified by our numerical analysis.
Our model possesses two parameters, which are the final black hole mass and the mass injection speed.

We find there are three cases of the dynamical D7-brane embeddings depending on these parameters. Two of them are analogous to static embeddings.
If the final black hole mass is sufficiently small, the brane never
intersects with the event horizon, and the embedding is the Minkowski
embedding with non-decaying oscillations. 
In the dual field theory, the quark condensate $c(V)$ oscillates around the
equilibrium value.
This is the sub-critical case in which the phase transition does not occur
since the final temperature is sufficiently low.
If the final mass is sufficiently large, the brane intersects with the horizon, and relaxes to the black hole embedding. 
This is the super-critical case in which the final temperature is
sufficiently high for the phase transition to occur and the mesons are melting.

We find that another kind of dynamical embedding appears when the final black hole mass is slightly smaller than the phase transition value.
There is no static BH embeddings in this regime.
Despite this fact, the brane dynamically touches the
horizon if the injection is sufficiently fast:
The rapid injection induces a large brane motion, and it drives the brane into the horizon.
Roughly speaking, a part of the brane falls into the black
hole because of inertia.
We call it the \textit{overeager case} 
since the brane is made ``overeager'' to plunge into the black hole under the influence of the dynamical environment.

\begin{figure}[tbp]
  \centering
  \subfigure[Sub-critical]
  {\includegraphics[scale=0.6]{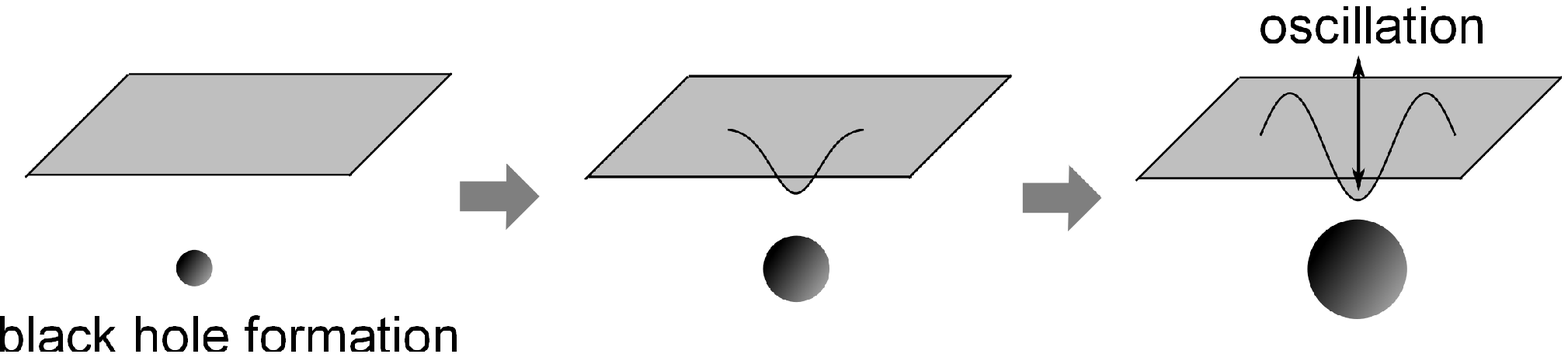}
\label{ponchi_mnk}
  }
  \subfigure[Super-critical]
  {\includegraphics[scale=0.6]{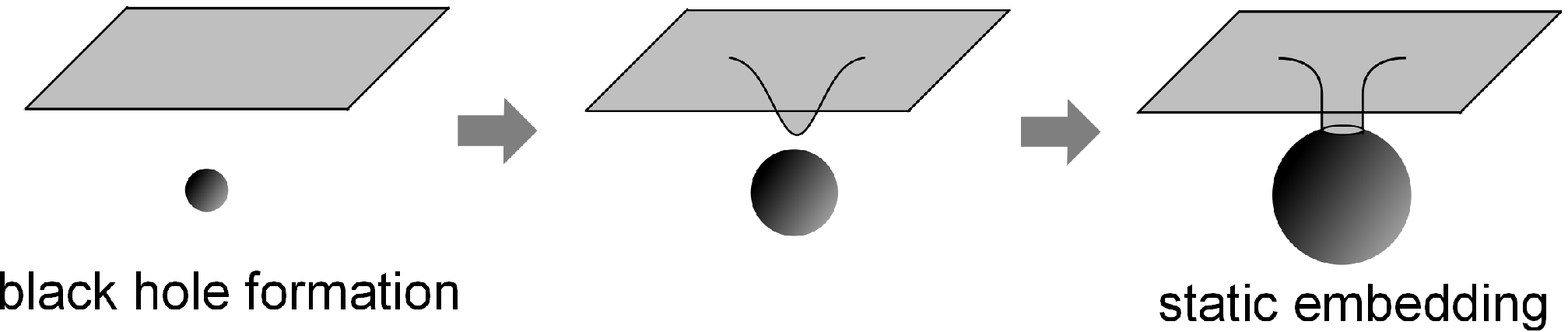}
\label{ponchi_bh}
  }
  \subfigure[Overeager]
  {\includegraphics[scale=0.7]{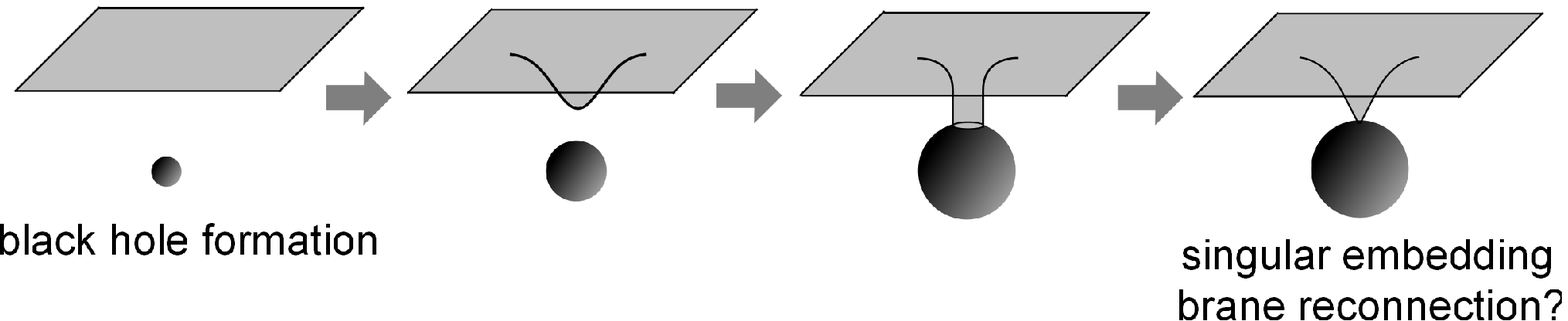}
\label{ponchi_isami}
  }
  \caption{
Schematic illustration of the brane dynamics.
\label{ponchis}
}
\end{figure}

After intersecting, the brane 
configuration on constant $V$ surfaces
moves in order to escape from the bulk black hole 
since the final equilibrium configuration in the future 
cannot
be the black hole embedding.
The brane configuration, however, tends to be singular within a finite time as the
intersection locus approaches the pole where the $S^3$ 
wrapped by the D7-brane
shrinks to zero size.
It is expected that
if stringy or quantum effects are taken into account there, the brane will reconnect near that locus, and then go back to the Minkowski embedding.
We show schematic illustration of this process in Fig.~\ref{ponchis}(c), together
with 
the sub-critical and super-critical cases
in Figs.~\ref{ponchis}(a) and \ref{ponchis}(b), respectively.
We argue that this phenomenon may be the gravity dual of quarks recombining into mesons in the boundary field theory.

The remaining of this paper is organized as follows. In Section~\ref{sec:static_embeddings}, we will review the static embeddings to fix notations. In Section~\ref{sec:dynamical_embeddings}, we describe the setup for our dynamical computation. Results of numerical computations are shown in Section~\ref{sec:results}. Section~\ref{sec:discussions} is devoted to discussions on the boundary theory interpretations and future directions. 
Technical details for numerical computations are provided in appendices.

\section{Static embeddings}
\label{sec:static_embeddings}

We start from reviewing the static embeddings of the D7-brane to the Schwarzschild-AdS$_5\times S^5$ spacetime in order to fix notations.
The metric of the Schwarzschild-AdS$_5\times S^5$ background can be given as
\begin{equation}
 ds^2_{10}=\frac{L^2}{z^2}\left[-f(z)dt^2 + \frac{dz^2}{f(z)} +d\vec{x}_3^2\right] +
  L^2 (d\phi^2+\cos^2\phi d\Omega_3^2 + \sin^2\phi d\psi^2)\ ,
\end{equation}
where $f(z)=1-r_\mathrm{h}^4z^4/L^8$, $d\vec{x}_3^2=dx_1^2+dx_2^2+dx_3^2$ and $L$ is
the AdS radius. The AdS boundary is at $z=0$, while the event horizon is at $z=L^2/r_\mathrm{h}$.
We use the same world-volume coordinates as the target space
coordinates themselves such that 
$(\sigma^0$, $\sigma^1$, $\cdots$, $\sigma^7)=(t,z,\vec{x}_3,\Omega_3)$.
The embedding of the D7-brane is specified by its position $(\phi, \, \psi)$ in the transverse space as a function of $z$: 
$\phi=\Phi(z)$ and $\psi=0$. Note that we can set $\psi=0$ without loss
of generality thanks to the $U(1)$-symmetry generated by
$\partial_\psi$. The induced metric on the D7-brane is given as
\begin{equation}
 L^{-2}h_{ab}d\sigma^a d\sigma^b= - \frac{f(z)}{z^2}dt^2
  + \left[\frac{1}{z^2f(z)} + \Phi'(z)^2\right] dz^2 
  + \frac{1}{z^2} d\vec{x}_3^2
  + \cos^2\Phi(z) d\Omega_3^2\ ,
\end{equation}
where ${}^\prime \equiv d/dz$. 

The embedding of the D7-brane is determined by the Dirac--Born--Infeld (DBI) action.
In the absence of the world-volume field strength, the action is
\begin{equation}
 S_{\mathrm{D}7} \propto \int d^8\sigma \sqrt{h}\ ,
\label{DBI}
\end{equation}
where $h = \det h_{ab}$. 
The equation of motion is a second-order ordinary differential equation for $\Phi(z)$,
\begin{equation}
\Phi''
-\frac{z}{2}(8f-zf')\Phi'{}^3
+ (3\tan\Phi) \Phi'{}^2
+\left(\frac{f'}{f}-\frac{3}{z}\right)\Phi'
+\frac{3\tan\Phi}{z^2f}
=0\ .
\label{Phieqst}
\end{equation}
It is also convenient to introduce new bulk coordinates $(w,\rho)=(L^2z^{-1}\sin\phi, L^2z^{-1}\cos\phi)$.
The new embedding function can then be a function of $\rho$, $w=W(\rho)$.
In practical numerical computations, we solve the equation for $W(\rho)$ obtained by rewriting Eq.~\eqref{Phieqst} in terms of $W(\rho)$.

The asymptotic behavior of the field at the AdS boundary gives information of the dual field theory operators.
Near the AdS boundary $\rho=\infty$, the asymptotic solution behaves as\footnote{
In the notation of Ref.~\cite{Mateos:2007vn}, $W$ is expanded as 
$W=2^{-1/2}r_\mathrm{h}\tilde m + 2^{-3/2}r_\mathrm{h}^3 \tilde c/\rho^2+\cdots$.
In this notation, the quark mass and condensate are
written as
$M_\mathrm{q}=\sqrt{\lambda}T\tilde m/2$ and 
$\langle\mathcal{O}_m\rangle=-\sqrt{\lambda}N_\mathrm{f}N_\mathrm{c}T^3\tilde c/8$, 
where $T\equiv r_\textrm{h}/(\pi L^2)$ is the Hawking temperature and
$\lambda$ is the 't Hooft coupling.
}
\begin{equation}
 W(\rho)=m+\frac{c}{\rho^2}+\cdots\ ,
\end{equation}
where the two integral constants $m$ and $c$ are related to the quark mass 
$M_q$
and condensate $\langle \mathcal{O}_m\rangle$ as~\cite{Kruczenski:2003uq,Babington:2003vm,Kirsch:2004km,Apreda:2005yz,Ghoroku:2005tf,Mateos:2006nu}
\begin{equation}
M_\mathrm{q}=\frac{m}{2\pi \ell_\mathrm{s}^2}\ ,\qquad
\langle \mathcal{O}_m\rangle = -\frac{N_\mathrm{f}
}{16\pi^4g_\mathrm{s} \ell_\mathrm{s}^6}c\ ,
\label{Mq}
\end{equation}
where $\ell_s$ is the string length and $g_s$ is the string coupling constant.
$\mathcal{O}_m$ is the quark bilinear associated with its supersymmetric counterparts (See Refs.~\cite{Mateos:2007vn,Kobayashi:2006sb} for further details).
Hereafter, we ignore the proportional constants 
and refer to $m$ and $c$ as the quark mass and the quark condensate for notational brevity.

In the bulk, we also need to impose boundary conditions at where the D7-brane terminates. The boundary conditions depend on the topology of the embedding, that is, whether the D7-brane intersects the black hole horizon or not.
When the brane intersects with the black hole,
we impose regularity at the horizon. The asymptotic solution is then obtained as
\begin{equation}
 W(\rho)=\sqrt{r_\mathrm{h}^2-\rho_\mathrm{H}^2}+\frac{\rho_\mathrm{H}\sqrt{r_\mathrm{h}^2-\rho_\mathrm{H}^2}}{\rho_\mathrm{H}^2+3r_\mathrm{h}^2}(\rho-\rho_\mathrm{H})+\cdots\ ,
\label{eq:BH_BC}
\end{equation}
where $\rho_\mathrm{H}$ ($0 < \rho_\mathrm{H} \le r_\mathrm{h}$) is the value of $\rho$ at the horizon found in the $(\rho, w)$-plane.
On the other hand, when the brane does not 
intersect with the black hole, the brane terminates at the point where $S^3$ wrapped by the brane 
shrinks to zero size at a pole of $S^5$: $\Phi=\pi/2$, corresponding to $\rho=0$.  
The regular asymptotic solution at $\rho=0$ is given by
\begin{equation}
 W(\rho)=W_\mathrm{P}-\frac{r_\mathrm{h}^4}{4W_\mathrm{P}^5}\rho^2 +\cdots\ ,
\label{eq:Min_BC}
\end{equation}
where $W_\mathrm{P} \equiv W(0) > r_\mathrm{h}$ is the value of $W$ at the pole.

\begin{figure}
\begin{center}
\includegraphics[scale=0.4]{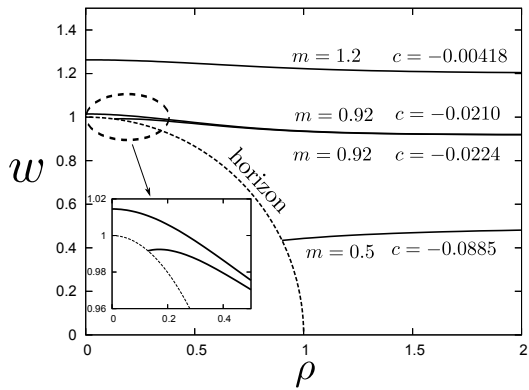}
\end{center}
\caption{
Minkowski and black hole embeddings of the D7-brane in the 
Schwarzschild-AdS$_5\times S^5$ spacetime. 
The vertical and horizontal axes are $w$ and $\rho$, respectively.
In the plot, the horizon radius is fixed
 as $r_\mathrm{h}=1$ 
using the scaling symmetry.
For $m\gtrsim 0.92$ and $m\lesssim 0.92$, we obtain the
 Minkowski and black hole embeddings, respectively.
For $m\simeq 0.92$, there exist both the Minkowski and black hole embeddings.
}
 \label{static}
\end{figure}

\begin{figure}
\begin{center}
\includegraphics[scale=0.5]{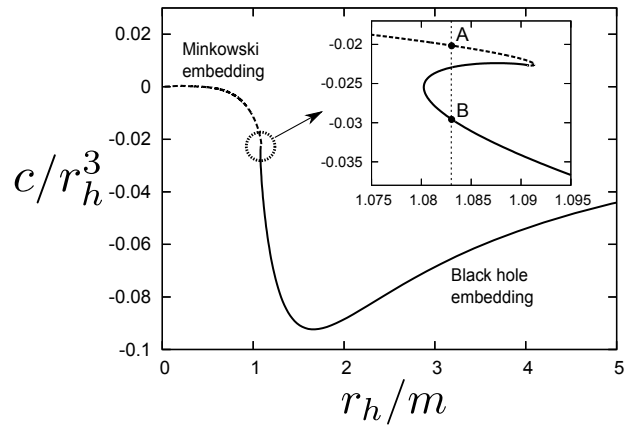}
\end{center}
\caption{
The quark condensate $c/r_\mathrm{h}^3$ against $r_\mathrm{h}/m$. 
The condensate is a multi-valued function in $1.0802<r_\mathrm{h}/m<1.0913$, 
and makes a finite jump between the points A and B by the
 first-order phase transition.
}
 \label{cm}
\end{figure}

Solving the equations of motion by using either the boundary condition \eqref{eq:BH_BC} or \eqref{eq:Min_BC} when the brane intersects with the black hole or not, respectively, we obtain a sequence of embeddings.
In Fig.~\ref{static}, we show the solutions for several parameter values.
For $m/r_\mathrm{h}\simeq 0.92$, we can see that both the black hole and Minkowski embeddings are allowed, and they are characterized by different values of the quark condensate $c$ for a given $m$. 
We plot $c/r_\mathrm{h}^3$ as a function of $r_\mathrm{h}/m$ in Fig.~\ref{cm}.
$c/r_\mathrm{h}^3$ becomes multivalued for $1.0802<r_\mathrm{h}/m<1.0913$. 
This implies that there is a first-order phase transition between the 
Minkowski and black hole embeddings, and it can be confirmed by comparing the free energy~\cite{Mateos:2006nu,Mateos:2007vn}.




\section{Dynamical embeddings}
\label{sec:dynamical_embeddings}

\subsection{Vaidya-AdS$_5$ spacetime}

In this section, we consider the very-far-from-equilibrium dynamics of the probe D7-brane in a dynamical
spacetime focusing on the thermalization process.
Thermalization in the boundary theory is realized in the gravity dual by the black hole
formation due to gravitational collapse in the bulk AdS space.
In this paper, we use the Vaidya-AdS spacetime as the
dynamical background with the black hole formation. 
Hereafter, we set units where the AdS radius is unity, $L=1$.
The Vaidya-AdS$_5$ metric is given by 
\begin{align}
g^{(5)}_{\mu\nu}dx^\mu dx^\nu
&=\frac{1}{z^2}[-F(V,z)dV^2 - 2dV dz +d\vec{x}_3^2]\ , \label{bg} \\
F(V,z)&=1-M(V)z^4\ ,
\end{align}
where the mass function $M(V)$ is a free function representing the
Bondi mass (density) of this spacetime. 
This metric is an exact solution of the following five-dimensional
Einstein equation with null dust, 
\begin{equation}
 G_{\mu\nu}-6g^{(5)}_{\mu\nu}= \frac{3z^3}{2}\frac{dM}{dV} k_\mu k_\nu\ ,
\end{equation}
where $k_\mu$ denotes the ingoing null vector defined by $k_\mu dx^\mu = - dV$.
The null dust injected from the AdS boundary infalls into the bulk, 
and then the event horizon is formed.
When the mass function $M(V)$ stops depending on $V$, namely, when the energy
injection ceases, the spacetime is
locally isometric to the Schwarzschild-AdS$_5$ with a planar horizon.

In our calculations, we chose a $C^1$ function for $M(V)$ as
\begin{equation}
M(V)=
\begin{cases}
0 & (V<0)\\
M_f \frac{1-\cos(\pi V/\Delta V)}{2} & (0 \le V \le \Delta V)\\
M_f & (\Delta V<V)
\end{cases},
\label{massfunc}
\end{equation}
where $M_f$ is the final mass of the spacetime, and $\Delta V$ determines  
the duration of the energy injection.
Note that the final radius of the event horizon, $r_\mathrm{h}$, is given by 
$M_f = r_\mathrm{h}^4$.
Uplifting the solution to $10$-dimensions, we obtain Vaidya-AdS$_5\times S^5$
spacetime:
\begin{equation}
 ds^2_{10}=\frac{1}{z^2}[-F(V,z)dV^2 - 2dV dz +d\vec{x}_3^2] + d\phi^2+\cos^2\phi d\Omega_3^2 + \sin^2\phi d\psi^2\ ,
\end{equation}
We consider the probe branes on this $10$-dimensional background geometry.
Figure~\ref{ponchi} gives schematic illustration of our setting, as explained in Sec.~\ref{subsec:setup}.

%

\subsection{Evolution equations of the D7-brane}

Having explained the background spacetime, we proceed to consider evolution
equations of the D$7$-brane to solve dynamical embeddings numerically.
Dynamics of a probe D-brane is described by the DBI
action. It has the diffeomorphism invariance with respect to
transformations of world-volume coordinates.
We should choose some useful coordinates, namely fix the gauge,
for solving the evolution equations numerically.
In order to derive equations of motion taking advantage of coordinate
transformations on the brane, here we would like to begin with the Polyakov type action
instead of the DBI action.
Of course, once we know the appropriate ansatz, we could simply start from the DBI action to derive the equations of motion. The result is the same as that obtained from the Polyakov type action since the computation is on-shell.

The Polyakov type action for the D7-brane in the absence of the world-volume field
strength is given by 
\begin{equation}
 S_\mathrm{P}=\int d^8\sigma \sqrt{\gamma}\,(
g_{\mu\nu}\gamma^{ab}\partial_a X^\mu \partial_b X^\nu
- \lambda
)\ ,
\label{Polac}
\end{equation}
where $X^\mu$ is the brane collective coordinate and $g_{\mu\nu}$ is the
metric in the target space. 
Note that $\lambda$ is an arbitrary non-zero constant.
Variating the action with respect to $X^\mu$ and $\gamma_{ab}$, we
obtain the equations of motion of the D7-brane as\footnote{
Using Eq.~\eqref{EOM2} and eliminating the auxiliary field
$\gamma_{ab}$ from the action~\eqref{Polac}, we obtain Dirac--Nambu--Goto 
action: $S\propto \int d^8\sigma\sqrt{h}$.
}
\begin{align}
&D^2X^\mu+\Gamma^\mu_{\rho\sigma} D_a X^\rho D^a X^\sigma=0\ ,\label{EOM1}\\
&\gamma_{ab}=6\lambda^{-1}g_{\mu\nu}D_a X^\mu D_b X^\nu\equiv 6\lambda^{-1}h_{ab}\label{EOM2}\ ,
\end{align}
where $D_a$ is the covariant derivative with respect to $\gamma_{ab}$,
and $\Gamma^\mu_{\rho\sigma}$ is the Christoffel symbol with respect to $g_{\mu\nu}$.
The second equation implies that $\gamma_{ab}$ can be regarded as
the induced metric on the brane up to a factor $6\lambda^{-1}$.
One finds that the wave equations~\eqref{EOM1} for $X^\mu(\sigma)$ on the
world-volume with the induced
metric $\gamma_{ab}$ are evolution equations, and
the equations~\eqref{EOM2} which determine components of the induced metric
are constraint equations. 
In the following, we will study the D7-brane embeddings on
the Vaidya-AdS background spacetimes solving Eqs.~\eqref{EOM1} and \eqref{EOM2}.

There are eight world-volume coordinates $\sigma^a$
($a=0,1,\cdots,7$) on the brane.
For six of them, we use the target space
coordinates themselves as
$(\sigma^2,\cdots,\sigma^7)=(\vec{x}_3,\Omega_3)$ because we assume that
the brane has 
the same symmetry as the background bulk spacetime
in those directions.
For the other two coordinates, 
we introduce the double null coordinates $(\sigma^0,\sigma^1)=(u,v)$, and then
the brane collective coordinates are parametrized as
\begin{equation}
V=V(u,v)\ ,\quad
z=Z(u,v)\ ,\quad
\phi=\Phi(u,v)\ ,\quad
\psi=0\ ,
\label{VZPpara}
\end{equation}
where we set $\psi=0$ without loss of generality because of the $U(1)$-symmetry generated by $\partial_\psi$. 
The $(u,v)$-coordinates are chosen so that the $\partial_u$ and
$\partial_v$ are null generators. Then, the brane induced metric is
written as
\begin{equation}
 h_{ab}d\sigma^a d\sigma^b
= - 2A(u,v)dudv+\frac{1}{Z(u,v)^2}d\vec{x}_3^2+\cos^2\Phi(u,v) d\Omega_3^2\ .
\label{ind}
\end{equation}
In the following calculation, it is convenient to use a new
variable 
$\Psi(u,v)\equiv \Phi(u,v)/Z(u,v)$ 
instead of $\Phi(u,v)$.
From the $(u,v)$-component of Eq.~\eqref{EOM2}, we can express $A(u,v)$ in
term of 
$(V,Z,\Psi)$ 
as 
\begin{equation}
 A=
\frac{1}{Z^2}[
FV_{,u}V_{,v}+Z_{,u}V_{,v}+Z_{,v}V_{,u}
]
- (Z\Psi)_{,u}(Z\Psi)_{,v}
\ .
\end{equation}
Note that there are still residual coordinate freedoms on the world-volume,
\begin{equation}
 \bar{u}=\bar{u}(u)\ ,\quad \bar{v}=\bar{v}(v)\ .
\label{resg}
\end{equation}
This residual gauge freedom will be fixed by boundary conditions and an
initial condition.
Substituting Eqs.~\eqref{VZPpara} and \eqref{ind} into Eq.~\eqref{EOM1},
we can obtain evolution equations for $V(u,v)$, $Z(u,v)$ and 
$\Psi(u,v)$.

The equations we will solve are summarized as follows.
The evolution equations for $V$, $Z$ and 
$\Psi$
are obtained as
\begin{align}
&V_{,uv}=
 \frac32 Z (Z\Psi)_{,u}(Z\Psi)_{,v}
+\frac32 \tan(Z\Psi)\{(Z\Psi)_{,u}V_{,v}+(Z\Psi)_{,v}V_{,u}\}
+\frac{1}{2}\left(F_{,Z}-\frac{5F}{Z}\right)V_{,u}V_{,v}
\ ,\label{EV1}
\\
&Z_{,uv}=
-\frac32 FZ (Z\Psi)_{,u}(Z\Psi)_{,v}
+ \frac32 \tan(Z\Psi)\{(Z\Psi)_{,u}Z_{,v}+(Z\Psi)_{,v}Z_{,u}\}
+\frac{5}{Z}Z_{,u}Z_{,v}
\notag\\
&\hspace{5cm}
- \frac{1}{2}\left(
 F_{,Z}-\frac{5F}{Z} 
\right)(FV_{,u}V_{,v}+V_{,u}Z_{,v}+V_{,v}Z_{,u})
- \frac{F_{,V}}{2}V_{,u}V_{,v}
%
\ ,\label{EV2}\\
%
%
&\Psi_{,uv}=
\frac32\left(\Psi F + \frac{\tan(Z\Psi)}{Z} \right)(Z\Psi)_{,u}(Z\Psi)_{,v}
+ \frac{1 - 3 Z \Psi \tan(Z\Psi) }{2Z^2}
\{(Z\Psi)_{,u}Z_{,v}+(Z\Psi)_{,v}Z_{,u}\}
\notag\\
&
\hspace{0.5cm}
+ \frac{\Psi}{2Z}\left(
 F_{,Z}-\frac{5F}{Z} + \frac{3\tan(Z\Psi)}{Z^2\Psi}
\right)
(FV_{,u}V_{,v}+V_{,u}Z_{,v}+V_{,v}Z_{,u})
- \frac{3\Psi}{Z^2}Z_{,u}Z_{,v}
+ \frac{\Psi F_{,V}}{2Z}V_{,u}V_{,v}
\ .
\label{EV3}
\end{align}
From the requirements that the $(u,u)$ and $(v,v)$-components of Eq.~\eqref{EOM2} vanish, we obtain the
constraint equations, 
\begin{align}
&C_1\equiv 
\frac{
\cos^3(Z\Psi)
}{Z^5}\left(
FV_{,v}^2+2Z_{,v}V_{,v}
-Z^2(Z\Psi)_{,v}^2
\right)
=0\ ,\label{Con1}\\
&C_2\equiv 
\frac{\cos^3(Z\Psi)}{Z^5}
\left(FV_{,u}^2+2Z_{,u}V_{,u}
- Z^2 (Z\Psi)_{,u}^2
\right)=0\ .
\label{Con2}
\end{align}
Note that the evolution equations guarantee the conservation of the constraints: 
\begin{equation}
\partial_u C_1=\partial_v C_2=0\ .
\end{equation}
Consequently, once we set an initial data satisfying the constraint
equations, 
we may solve only the evolution equations as the constraints will be kept satisfied.
The details of the
numerical method to solve these equations are summarized in Appendix~\ref{sec:num}.

Eliminating $u$ and $v$, we can regard 
$\Psi$
as
a function of $Z$ and $V$. 
Near the AdS boundary $Z=0$, 
$\Psi(V,Z)$ 
is expanded as
\begin{equation}
\Psi = m + \left(
c(V) + \frac{m^3}{6}
\right)Z^2
+\cdots\ .
\end{equation}
The constant $m$ and the function $c(V)$ are related to the quark mass
and condensate as in Eqs.~\eqref{Mq}.
Solving the evolution equations, we can determine
the time dependence of the quark condensate $c(V)$.

\subsection{Boundary conditions and initial data}
\label{sec:BCandID}

In general, two timelike boundaries appear on the 
world-volume of the brane: 
one is the AdS boundary $Z=0$, and the other is the pole
$\Phi=\pi/2$ at which the radius of $S^3$ wrapped by the D7-brane
shrinks to zero.
When a numerical domain to be solved contains these boundaries, we need to impose
boundary conditions there.
Since, however, the evolution equations become singular at each
boundary, we should fix the location of the boundary in the world-volume $(u, \, v)$-coordinates
for numerical convenience.

As mentioned previously, we have the residual gauge freedom, which is a transformation from
the original null coordinates to other ones~\eqref{resg}.
This is nothing but a conformal transformation between two-dimensional
flat spacetimes.
Using it,
one can fix the location of the AdS boundary on the
world-volume coordinates as $u=v$, i.e. $Z|_{u=v} = 0$, or that of the
pole as $u=v+\pi/2$, i.e. $\Phi|_{u=v+\pi/2} = \pi/2$, or both.
They become coordinate conditions on the world-volume domain.


Note that fixing the residual gauge is equivalent to choosing a certain conformal factor, which behaves as a harmonic function obeying a $(1+1)$-dimensional free massless field equation.%
\footnote{
We denote a conformal transformation as 
$\bar\gamma_{ab} = e^{\omega}\gamma_{ab}$.
In two dimensions, the scalar curvature is 
$\bar R=e^{-\omega}(R-\nabla^2\omega)$, where $\nabla$ is the covariant
derivative with respect to $\gamma_{ab}$.
Now, since both $\bar\gamma_{ab}$ and $\gamma_{ab}$ are flat, namely
$R=0$ and $\bar{R}=0$, the
conformal factor must be determined by a harmonic function $\omega$
satisfying $\nabla^2 \omega = 0$.
}
Hence, if we give a set of coordinate conditions at the boundaries and the initial surface,
they will provide necessary boundary and initial conditions for the conformal factor.
Those conditions will fix the residual gauge completely, 
and the null-coordinate system will be automatically determined on the computational domain.

\subsubsection{Boundary conditions at the AdS boundary}
Now, we derive boundary conditions at the AdS boundary.
Since we will focus on observables there, the AdS boundary should be
always contained in our computational domain and located at $u=v$
throughout our calculations.
The boundary conditions for $Z$ and $W$ are trivial: 
$Z|_{u=v}=0$ and 
$\Psi|_{u=v}=m$.
We choose the quark mass as time-independent, namely $m$
is constant.
Remaining boundary conditions on the AdS boundary can be derived by requiring
regularities of the evolution equations at $Z=0$.
We expand the variables near the boundary as $V=V_0(v)+V_1(v)(u-v)+\cdots$, 
$Z=Z_1(v)(u-v)+Z_2(v)(u-v)^2+\cdots$ and 
$\Psi=m+\Psi_1(v)(u-v)+\cdots$. 
Substituting them into the evolution 
equations~\eqref{EV1}--\eqref{EV3}, we
obtain the asymptotic solution as 
\begin{equation}
\begin{split}
&V=V_0(v)+\mathcal{O}((u-v)^3)\ ,\quad
\Psi=m+\mathcal{O}((u-v)^2)
\ ,\\
&Z=\frac{\dot{V}_0}{2}(u-v)+\frac{\ddot{V}_0}{4}(u-v)^2+\mathcal{O}((u-v)^3)\ ,
\end{split} 
\end{equation}
where ${}^\cdot\equiv d/dv$.
They give 
\begin{equation}
 \dot{V}_0(v)=2Z_{,u}|_{u=v}\ ,\quad Z_{,uv}|_{u=v}=0\ .
\label{AdScond}
\end{equation}
Hence, integrating $Z_{,u}|_{u=v}$ along the AdS boundary, we can
determine the time evolution of $V|_{u=v}$.
(The second equation is used to determine $Z_{,u}$ at the boundary. For
details, see appendix~\ref{sec:num}.)
Note that while the regularities yield $V_{,u}|_{u=v}=0$ or
$V_{,v}|_{u=v}=0$, we have taken $V_{,u}|_{u=v}=0$ because the time
direction in our setup leads to $V_{,v}|_{u=v}>0$.
We can confirm that $V_{,u}|_{u=v}=0$ and $V_{,v}+2Z_{,v}|_{u=v}=0$ (it
is equivalent to the first equation of \eqref{AdScond}) are
consistent with the constraint equations at the AdS boundary.

\subsubsection{Boundary conditions at the pole}
Next, we consider the boundary conditions at the pole, at which the
radius of $S^3$ wrapped by the D7-brane shrinks to zero.
The brane may intersect with the event horizon in our calculations,
and in that case
we do not need to impose the boundary conditions at the pole since the
AdS boundary is not in the causal future of the pole on the brane.
Strictly speaking, when our computational domain does not contain this
boundary, we do not need any boundary conditions there.
If we need the boundary conditions, we will locate the pole $\Phi=\pi/2$ at
$u=v+\pi/2$ by using the residual gauge.
Hence, one of the boundary conditions is 
$Z\Psi|_{u=v+\pi/2} = \pi/2$.
The others are given by regularities of the evolution equations at
$\Phi=\pi/2$ as 
\begin{equation}
 Z_{,u} = Z_{,v}\ , \quad V_{,u} = V_{,v}\ ,
\end{equation}
at $u=v+\pi/2$.
They are merely Neumann boundary conditions at the pole.

\subsubsection{Initial data}
Finally, we explain the initial data for our calculations.
Before the energy injection $V<0$, 
the background spacetime is pure AdS, and the brane is static there.
We have the exact solution of the static embedding given by 
$Z=m^{-1}\sin\Phi$, i.e., 
$\Psi=Z^{-1}\arcsin(mZ)$.
Solving the constraint equations \eqref{Con1} and
\eqref{Con2}, we obtain
\begin{equation}
\begin{split}
&V(u,v)=m^{-1}(\phi(u)+\phi(v)-\sin(\phi(u)-\phi(v))) + V_\mathrm{ini}\ ,\\
&Z(u,v)=m^{-1}\sin(\phi(u)-\phi(v))\ ,\quad
\Psi(u,v)=\frac{m(\phi(u)-\phi(v))}{\sin(\phi(u)-\phi(v))}
\ ,
\end{split}
\end{equation}
where $\phi$ is a free function corresponding to the residual coordinate
freedom, and $V_\mathrm{ini}$ is an integration constant.
At the initial surface $v=0$, the solutions become
\begin{equation}
V(u,0)=m^{-1}(\phi(u)-\sin\phi(u)) + V_\mathrm{ini}\ ,\quad
Z(u,0)=m^{-1}\sin\phi(u)\ ,\quad
\Psi(u,0)=\frac{m\phi(u)}{\sin\phi(u)}
\ ,
\label{initial_data}
\end{equation}
where we have set $\phi(0)=0$, and then $V(0,0)=V_\mathrm{ini}<0$
is the initial time at the AdS boundary.

The residual gauge $\phi(u)$ corresponds to the coordinate freedom for $u$ on the initial surface, and then we can choose an appropriate function for $\phi(u)$ arbitrarily.
When the brane does not intersect with the event horizon,
we can simply choose $\phi(u)=u$. 
In this case, the pole $\Phi=\pi/2$ can be fixed at $u=v+\pi/2$.


If the brane intersects with the event horizon at the initial surface,
or if the brane is close to the event horizon, the above choice 
$\phi(u)=u$ is not good.
The reason is as follows.
Each $u$-constant line represents an out-going null geodesic on the
brane world-volume.
Roughly speaking, $\delta \phi \simeq \phi(u + \delta u) - \phi(u)$ is
a proper distance between two null geodesics described by $u+\delta u$
and $u$ at the initial surface.
If these out-going null geodesics pass through the vicinity of the event
horizon towards the AdS boundary, the interval of each arrival time at
the AdS boundary,
$\delta V \simeq V(u+\delta u, u+\delta u) - V(u,u)$, will become very
large compared to the proper distance $\delta\phi$ at the initial surface.%
\footnote{Rough estimation is 
$\delta V / \delta\phi \sim \kappa^{-1}(\phi(u_\mathrm{EH})-\phi(u))^{-1}$, 
where $\kappa$ is a surface gravity and $u_\mathrm{EH}$ is the coordinate
value describing the event horizon.}
This behavior essentially originates from the gravitational red-shift effect near the horizon.
As a result, if we choose $\phi(u)=u$, $V$ blows up rapidly at late time
and then the numerical calculations easily break down.

To avoid this problem, we
fine-tune $\phi (u)$ so that $V$ and $v$ are
synchronized at the AdS boundary, i.e., $V_{,v}|_{u=v}=1$ 
(or $V|_{u=v}=v+V_\mathrm{ini}$).
This implies that $d\phi/du$ becomes exponentially small in the vicinity
of the event horizon.
In Appendix~\ref{sec:num},
we explain how to fine-tune $\phi(u)$ in our numerical
calculations.

Note that the condition $V_{,v}|_{u=v}=1$ is 
another boundary condition at the AdS boundary imposed in addition to the one that fixes the boundary position at $u=v$.
This means that we do not fix the location of the pole $\Phi=\pi/2$ on the
world-volume coordinates.
However, this is not a problem because the pole is not contained in our
computational domain and then we do not need to impose the boundary conditions there.%

\section{Results}
\label{sec:results}
 
In this section, we show results of our numerical calculations.
We use units in which
$m=1$.\footnote{
In our system, there is a scaling symmetry:
$Z\to k Z$, $V\to kV$, $\Psi\to\Psi/k$, $\vec{x}_3\to k\vec{x}_3$, 
$M_f\to M_f/k^4$, $\Delta V\to k\Delta V$, 
$m\to m/k$, and $c\to c/k^3$,
where $k$ is a constant. Using the scaling symmetry, we can set $m=1$
without loss of generality.
}
The parameters are the final mass of the black
hole (in other words, the final temperature) and the duration of the
energy injection, which are respectively described by $r_\mathrm{h}$ and $\Delta V$.

\subsection{Sub-critical case}
\label{subsec:subcri}
 

In this subsection, we show numerical results for the small
final mass case in which the D$7$-brane does not touch the horizon and
remains in the Minkowski embedding after the black hole formation. 
This is the sub-critical case in which 
the final temperature is sufficiently low and the phase transition does not occur.
  Since there is no dissipation 
when the brane is in
  the Minkowski embedding, excitations on the brane caused by
  dynamics of the bulk geometry 
will remain without decay.

\begin{figure}[tbp]
   \begin{center}
    \includegraphics[scale=0.8,clip]{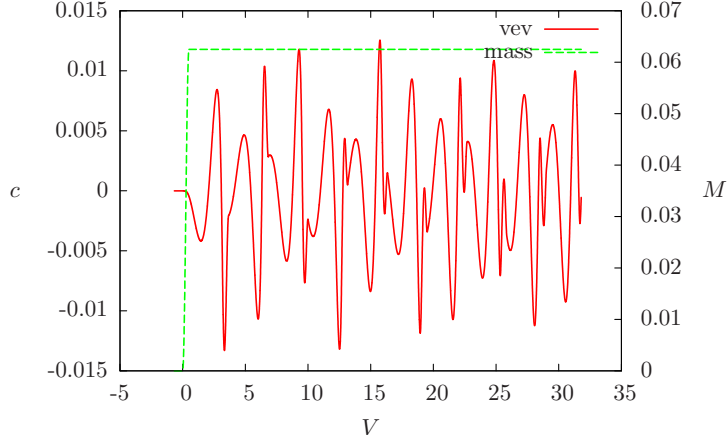}
    \caption{The quark condensate $c$ and the BH mass $M$ plotted as a function of the boundary time $V$ for energy injection duration
    $\Delta V = 0.5$. The horizon radius in the final state is
   $r_\mathrm{h} = 0.5$, for which the equilibrium value of $c$ 
    is $c_\mathrm{eq} = - 4.07 \times 10^{-5}$.}
    \label{fig:vev_zh2d05}
   \end{center}
  \end{figure}

\begin{figure}[tbp]
   \begin{center}
    \includegraphics[scale=0.8,clip]{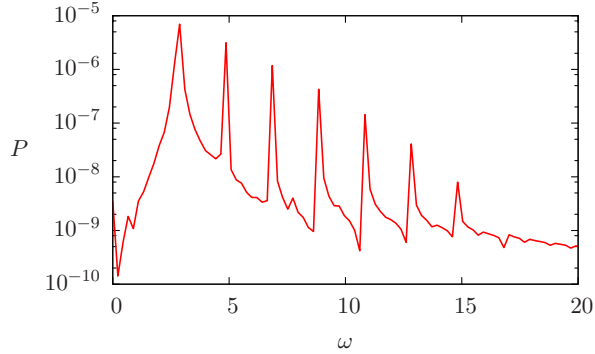}
    \caption{Power spectrum $P(\omega)$ of $c(V) - c_\mathrm{eq}$ for the same
    parameters as in Fig.~\ref{fig:vev_zh2d05}.
    It is obvious that the excitations have discrete spectrum, namely
    normal modes.
   The first peak in the spectrum is located at $\omega=2.8$, 
   and it corresponds to the lowest normal mode of the D7-brane oscillations.
    }
    \label{fig:ps_zh2d05}
   \end{center}
  \end{figure}

  Figure~\ref{fig:vev_zh2d05} shows the time evolution of the quark condensate
  $c$ and the Bondi mass $M$ with respect to the boundary time $V$.
  Note that all of these quantities are well-defined at the AdS
  boundary and are directly related with observables in the boundary theory.
  The Bondi mass $M$ corresponds to the energy density of the CFT
  (gluon) fluid.
  We can recognize seemingly periodic oscillations after the energy injection.
  These correspond to excitations of stable mesons, namely normal
  modes of the brane fluctuations.
  Indeed, the power spectrum of $c(V)$ shown in Fig.~\ref{fig:ps_zh2d05} obviously indicates
  that the excitations have discrete spectrum,
  similarly to the case of
  linear perturbations on the static Minkowski embedding.
  Note that we have solved non-linear evolution equations of the
  brane rather than linear perturbations around an equilibrium brane
  configuration, 
and the discrete spectrum is a nonlinear version of the normal mode spectrum for linear fluctuations.

\begin{figure}[tbp]
   \centering
   \subfigure[$c(V)$ for $M(V)$ of Eq.~(\ref{massfunc})]
   {\includegraphics[width=8cm,clip]{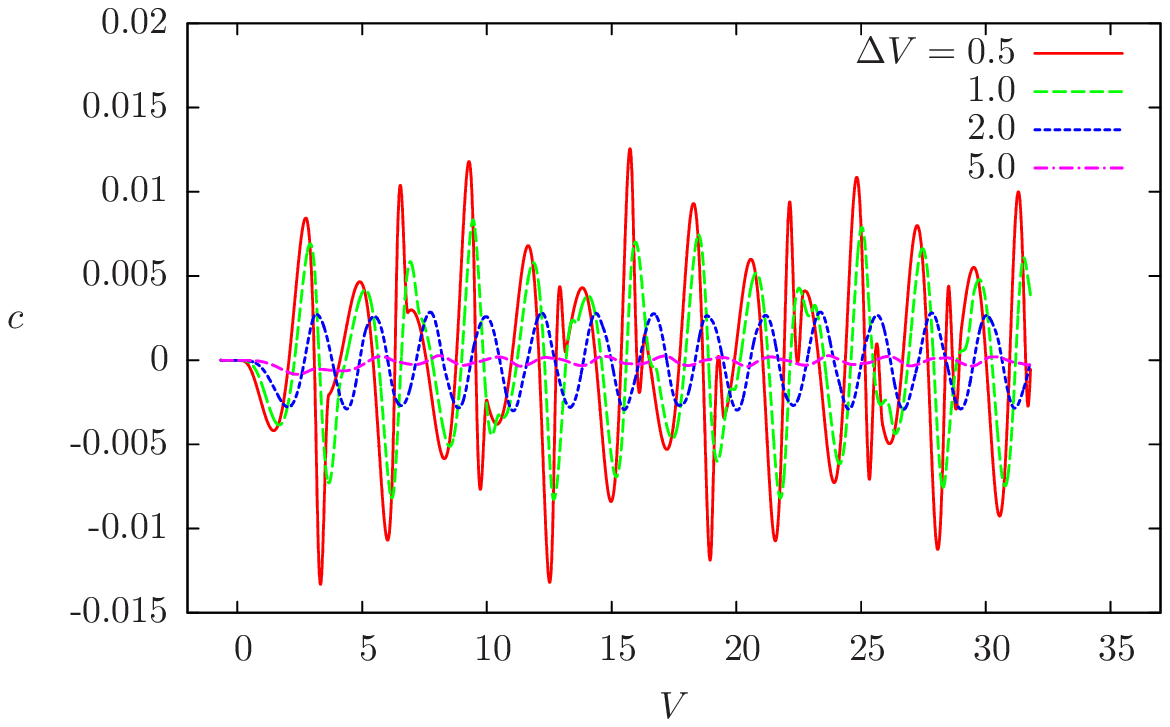}
  \label{Fig:vev_dVpanel}
   }
   \subfigure[$P_\text{tot}(\Delta V)$]
   {
  \includegraphics[width=8cm,clip]{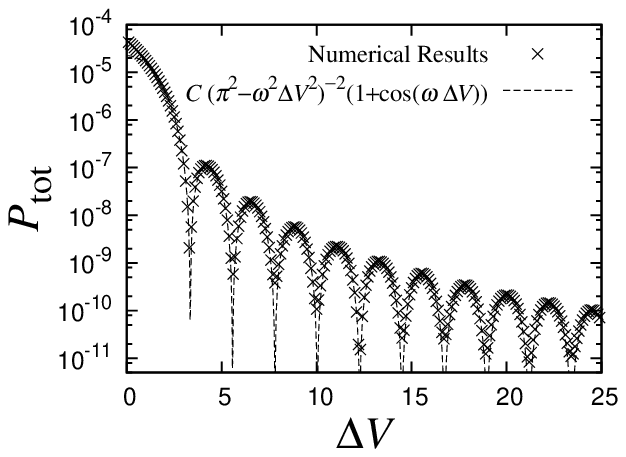}
  \label{Fig:Ptot}
   }
   \caption{
Panel~(a): Time evolution of the condensate $c$ for various injection
   duration $\Delta V$.
   The horizon radius at the final state is
   $r_\mathrm{h} = 0.5$ and the equilibrium value of $c(V)-c_\text{eq}$ 
    is $c_\mathrm{eq} = - 4.07 \times 10^{-5}$.
  Panel~(b):
   Total power $P_\mathrm{tot}(\Delta V)$ of the condensate $c(V)$.
}
   \label{fig:vev_dV}
  \end{figure}

  Figure~\ref{Fig:vev_dVpanel} shows time evolutions of $c(V)$ 
  for various $\Delta V$.
  It turns out that the shorter the injection duration $\Delta V$
  is, the harder the condensate $c(V)$ oscillates.
  This implies that the larger non-adiabaticity is, the larger the
  excitations on the brane are.
  To study the fluctuations in more detail, 
  we show the total power of fluctuations of the condensate,
  $P_\mathrm{tot}(\Delta V) \equiv \lim_{V_1\to \infty}\frac{1}{V_1-V_0}\int_{V_0}^{V_1} (c(V) - c_\mathrm{eq})^2 dV$,
  in Fig.~\ref{Fig:Ptot}.
  We calculated $c(V)$ for $0.1\leq \Delta V \leq 25$ setting the final horizon radius to $r_\text{h}=0.5$, and obtained $P_\text{tot}(\Delta V)$ from $c(V)$ using $V_0=30$ and $V_1=50$. We confirmed that the numerical results are almost independent of $V_1$ as long as we keep it larger than $50$.
   The numerical results imply that $P_\text{tot}(\Delta V)$ shows a damped oscillation whose amplitude behaves as $\Delta V^{-4}$, and appears to converge into the finite maximum value in the $\Delta V\to 0$ limit.

Before moving to the next subsection, we make some comments on the relationship of these results to a harmonic oscillator model.
We find that features of $P_\text{tot}(\Delta V)$ can be well reproduced by a harmonic oscillator with an external force described by
$\frac{d^2x(V)}{dV^2} + \omega^2 x(V) = f(V)$, if we take $M(V)$ of Eq.~(\ref{massfunc}) as the external force $f(V)$.
The total oscillation power in the final state of this model is given by
$P_\text{tot}(\Delta V) = \frac{C(1+\cos(\omega \Delta V))}{(\pi^2 - \omega^2 \Delta V^2)^2}$,
where $C$ is a normalization constant independent of $\Delta V$.%
\footnote{
When $f(V)$ is set to $M(V)$ of Eq.~(\ref{massfunc}) and the initial state is set to $x(0)=0=\frac{dx}{dV}(0)$, 
we can show that the total power at late time ($V>\Delta V$) is given by
$\langle(x-x_0)^2\rangle = \frac{\pi^4 x_0^2}{4}\frac{1+\cos(\omega \Delta V)}{(\pi^2 - \omega^2 \Delta V^2)^2}$,
where $x_0\equiv M_f/\omega^2$ is the equilibrium point of the oscillator at late time 
and $\langle A(V)\rangle$ is the average of a function $A(V)$ over a time period sufficiently longer than $\omega^{-1}$.
When $f(V)$ is changed into that of Eq.~(\ref{massfunc_linear}),
the total power at late time changes into
$\langle(x-x_0)^2\rangle = \frac{2x_0^2}{\omega^2\Delta V^2}\sin^2\frac{\omega\Delta V}{2}$.
}
 Fitting this analytic formula to the numerical data, we obtain $\omega = 2.8$. 
This value coincides with the lowest normal mode frequency of the brane oscillations (see Fig.~\ref{fig:ps_zh2d05}).

\begin{figure}[htbp]
\centering
   \includegraphics[width=8.5cm,clip]{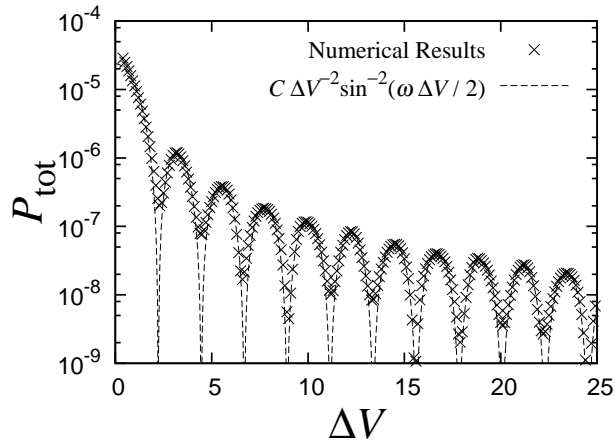}
\caption{Total power of the condensate for $M(V)$ given by
 Eq.~(\ref{massfunc_linear}) for the final horizon radius
 $r_\text{h}=0.5$.
$P_\text{tot}(\Delta V)$ shows a damped oscillation that decays as $\Delta V^{-2}$.}
  \label{Fig:Ptot_linM}
\end{figure}

Precise behavior of $P_\text{tot}(\Delta V)$ depends on the function form of $M(V)$.
As an example, in Fig.~\ref{Fig:Ptot_linM} we show $P_\text{tot}(\Delta V)$ for $M(V)$ given by
\begin{equation}
M(V)=
\begin{cases}
0 & (V<0)\\
M_f \frac{V}{\Delta V}
& (0 \le V \le \Delta V)\\
M_f & (\Delta V<V)
\end{cases}.
\label{massfunc_linear}
\end{equation}
For this $M(V)$, $P_\text{tot}(\Delta V)$ shows a damped oscillation that decays as $\Delta V^{-2}$.
Similarly to the previous case, 
this numerical result is well reproduced by a harmonic oscillator model:
When $f(V)$ is set to $M(V)$ of Eq.~(\ref{massfunc_linear}), the model gives
$P_\text{tot}(\Delta V)= \frac{C}{\Delta V^2}\sin^2\frac{\omega\Delta V}{2}$,
where the constant $C$ is independent of $\Delta V$.
Fitting to the numerical data gives $\omega=2.8$, 
and this value coincides with the lowest normal mode frequency.

These results suggest that some aspects of the time dependence of the D7-brane and the condensate $c(V)$ can be captured by the harmonic oscillator model with the external force $f(V)=M(V)$.
For example, the right-hand sides of the equations of motion (\ref{EV1})--(\ref{EV3}) contain terms proportional to $M(V)$, and those terms may be roughly identified with the external force applied to the D7-brane. 
It may be interesting to pursue these ideas in more detail.
Also, 
it may be interesting to study further on behaviors in the $\Delta V\to 0$ limit to
give an estimate on the largest amplitude realized by a rapid energy injection.
Our numerical results and the analytic results based on the harmonic oscillator model 
in this context
suggest that $P_\text{tot}(\Delta V)$ converges to finite constants in the $\Delta V\to 0$ limit.
It may be interesting to generalize results of Ref.~\cite{Buchel:2013gba}, 
which focuses on abrupt quenches in the holographic setup,
to the D3/D7 system and compare them with our results.

  \subsection{Super-critical case}
  \label{subsec:supcri}

In this subsection, we show numerical results for the large
final mass case in which the D7-brane is drawn into the event horizon
and its configuration changes from the Minkowski embedding to the BH
embedding. 
This is the super-critical case in which the final temperature is sufficiently
high for the phase transition to occur.

\begin{figure}[tbp]
   \begin{center}
    \includegraphics[scale=0.8,clip]{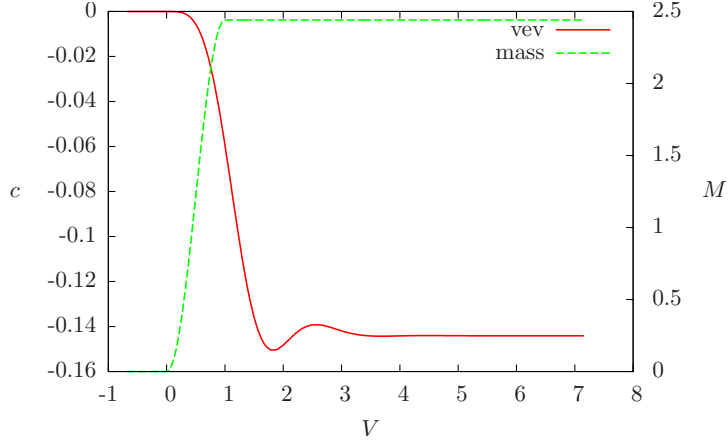}
    \caption{The quark condensate and BH mass for $\Delta V = 1.0$.
    The horizon radius in final state is $r_\mathrm{h} = 1.25$, for which
    equilibrium value of $c$ is $c_\mathrm{eq} = - 0.144$.}
    \label{fig:vev_zh08d1}
   \end{center}
  \end{figure}

  In Fig.~\ref{fig:vev_zh08d1}, we show a typical behavior of the quark
  condensate $c$ and the black hole mass $M$ in 
  this case, setting the energy injection duration to $\Delta V=1$.
  The quark condensate $c$ gradually evolves as the black hole mass increases due to the injection, and eventually settles down to the final value by $V\sim 4$. This final value coincides with $c$ of the static BH embedding for the same bulk black hole mass.
  It turns out that a quasi-normal mode is excited in the ring-down phase.
  This fact implies that the phase transition to the BH phase has
  taken place and the brane comes to equilibrium.
  
\begin{figure}[tbp]
   \begin{center}
    \includegraphics[scale=0.8,clip]{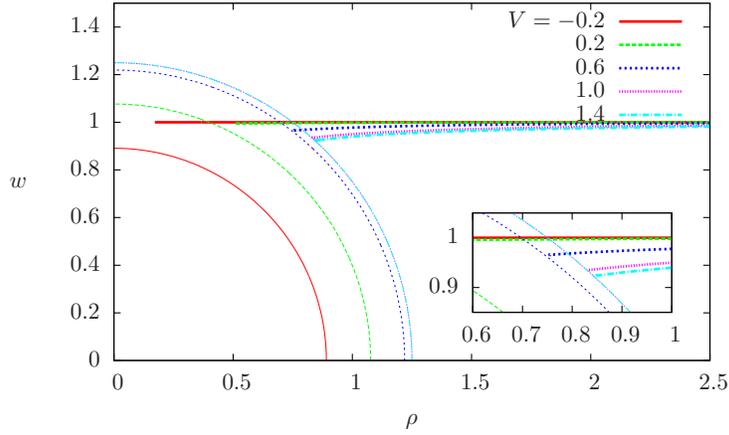}
    \caption{Snapshots of embeddings of the D$7$-brane in $(w,\rho)$ on
    time slices characterized by constant $V$. Circles are the
    event horizon at each time $V$. 
    The horizon radius at final state is $r_\mathrm{h} = 1.25$.
   For $V=-0.2$ and $0.2$ cases, the parts of the curves near the pole are not shown since they are not included in our numerical calculation domain.
}
    \label{fig:snapshot_BH-emb}
   \end{center}
  \end{figure}

In Fig.~\ref{fig:snapshot_BH-emb}, to focus on the brane
world-volume in the bulk, we show snapshots of the
brane configuration on time slices characterized by $V=\text{constant}$.
Note that we define the bulk coordinates on the time slices as
$(w,\rho)=(z^{-1}\sin\phi, z^{-1}\cos\phi)$.
The radius of the event horizon is evolving larger, and we can see that the
  brane eventually intersects with the horizon.
  At late time the configuration of the brane will coincide with the
  static BH embedding.

  When we consider dynamical phenomena in the bulk, 
  especially when an event horizon exists, 
  we should pay careful attention to the causality.
  In the static case, two meson phases of the boundary field theory can be distinguished by seeing
  whether the brane configuration at any moment is intersecting with the
  horizon or not.
  However, such a naive definition 
  can not be used in dynamical cases
  by the following reason.
  A temporal configuration of the brane, which is the intersection of the brane world-volume and a certain
  time slice of the bulk, depends on the given time slice.
  Since there is no definitely preferred time slice in general spacetime,
  we cannot uniquely identify 
  the phase of the brane embedding
  according to the temporal configuration at that time.
  To see it, let us consider the following example for
  time slices determined by $V=\text{const}.$
  In the current case, the event horizon does exist even before the injection starts and when the bulk spacetime is still locally pure AdS.
  The brane can enter the horizon within this early time domain.
  It means that the brane can intersect with the event horizon even at $V<0$
  (before the injection starts!).
   However, the brane configuration in that time domain is given by the exact solution in the pure AdS, 
  and it will be obviously regarded as the Minkowski embedding if static.


In the spirit of the gauge/gravity duality, 
to determine the phase of the brane embedding, we should look at the
quark condensate $c$
rather than the temporal configuration of the brane on a certain time slice.
If we consider the causality in the bulk further, we notice that observers outside the black hole (including those on the AdS boundary) cannot know whether the brane intersects the horizon or not, 
since the intersection locus lies on the event horizon and the exterior observers can never obtain information on that locus.
This implies that an attempt to determine the phase based on the
temporal brane configuration requires information inaccessible from the boundary.
On the other hand, the behavior of the quark condensate $c$ is defined only by the information outside the horizon, since this quantity is governed by the brane equations of motion that respect the bulk causality.
  Indeed, using the quark condensate $c$, we can examine the phase of the brane embedding without any problem, as shown in Fig~\ref{fig:vev_zh08d1}.
From the viewpoint of the boundary, the phase transition takes place
when the brane is drawn into the vicinity of the black hole 
(namely, 
when the gravitational red-shift at the brane becomes strong)
rather than
the inside of that.

  \subsection{Overeager case}
  \label{subsec:oe}

\begin{figure}[tbp]
  \begin{center}
  \includegraphics[scale=0.8,clip]{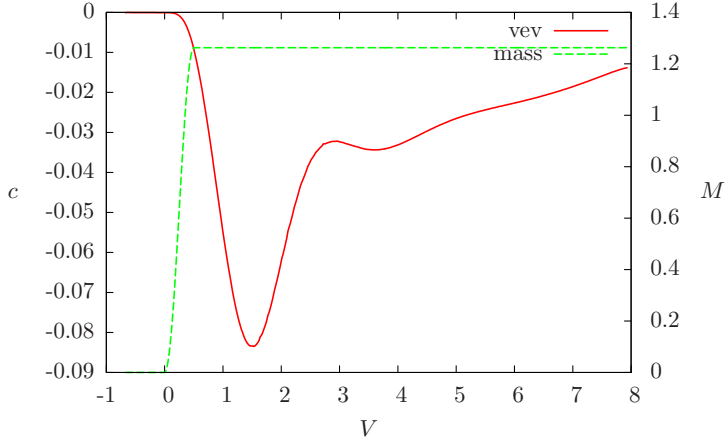}
  \caption{The quark condensate and the BH mass for duration of injection $\Delta V = 0.5$ 
  (rapid injection). 
  The horizon radius at final state is $r_\mathrm{h} = 1.06$, 
  for which only the Minkowski embedding with $c = - 2.00\times 10^{-2}$
  exists as a static solution.
  }
    \label{fig:vev_overeager}
  \end{center}
  \end{figure}

\begin{figure}[tbp]
   \begin{center}
    \includegraphics[scale=0.8,clip]{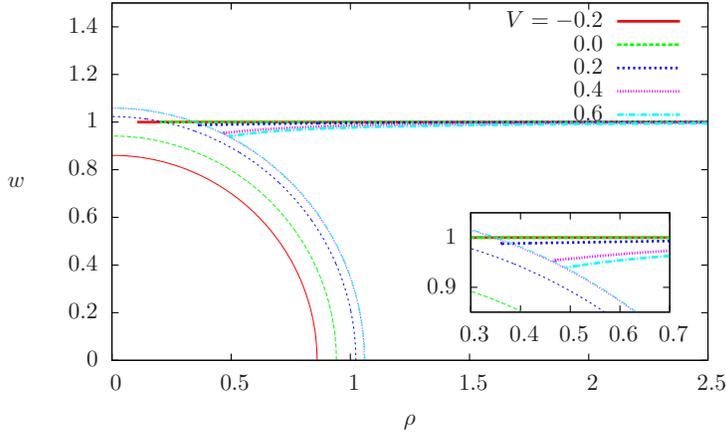}
    \caption{Snapshots of early-time evolution of embeddings for $V \le 0.6$. 
    The circles are the event horizon at each time $V$. 
    The brane has fallen through the event horizon.
    }
    \label{fig:snapshot_early}
   \end{center}
  \end{figure}

\begin{figure}[tbp]
   \begin{center}
    \includegraphics[scale=0.8,clip]{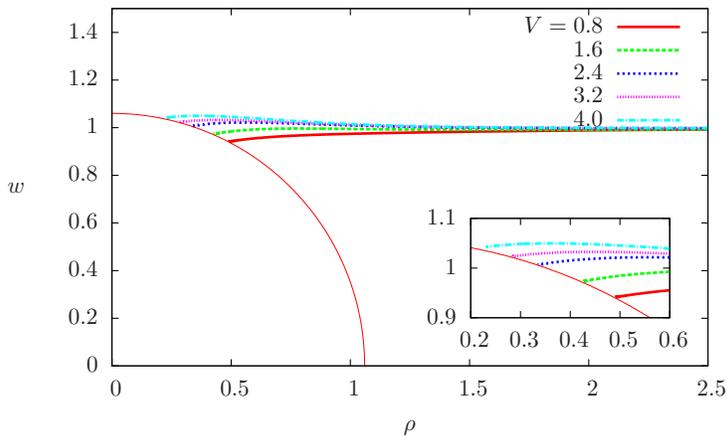}
    \caption{
    Snapshots of late-time evolution of embeddings for 
    $V \ge 0.8$. The circle is the event horizon at final state. 
    While the spacetime has settled down to the final state, the brane is
    still evolving gradually.
    }
    \label{fig:snapshot_late}
   \end{center}
  \end{figure}

  As a new significant feature in the dynamical evolution, we find that the brane
  can fall into the event horizon even in a parameter region where
  any equilibrium BH embedding does not exist.
  We would like to refer to this situation as the \textit{overeager case}, as explained in Sec.~\ref{subsec:summary}.
  Figure~\ref{fig:vev_overeager} shows the time evolution of quark
  condensate $c$ and the black hole mass $M$ in such an overeager case. 
  The time evolution of $c(V)$ markedly differs from
  the previous cases:
  There are neither periodic oscillations nor relaxation to the
  equilibrium value.
  Also, we show snapshots of the brane configuration
  at early time ($V \le 0.6$) and late time ($V \ge 0.8$) in
  Figs.~\ref{fig:snapshot_early} and \ref{fig:snapshot_late},
  respectively.
  We take $r_\mathrm{h}=1.06$ as the final radius of the
  black hole, for which only the Minkowski embedding exists as the
  static brane configuration (see Fig.~\ref{cm}).
  However, the brane configurations in Figs.~\ref{fig:snapshot_early}
  and \ref{fig:snapshot_late} indicate that the brane intersects with the event horizon.
  This fact is reflected in the non-periodic behavior of $c$.
  We also notice that the brane keeps moving even after the bulk spacetime has settled down to the final state.

  We can understand this bulk phenomenon in the following way.
  It is not forbidden that a part of the brane dynamically intersects with the event horizon even if any static BH embedding does not exist.%
\footnote{\label{footnote:DV0}
In a special case, this statement can be easily proved as follows.
  When $\Delta V$ is infinitesimal, namely for the fastest injection, 
  the radius of the event horizon becomes $r=r_\mathrm{h}$ just at $V=0$ and it will be the
  final radius.
  On the other hand, if the quark mass is $m<r_\mathrm{h}$, the tip of the static brane is located at $r=m\,(<r_\mathrm{h})$ before the injection $V<0$.
This implies that the portion of the brane near the tip resides within the event horizon at $V=0$.
  Since there is no equilibrium BH embedding for 
  $1<r_\mathrm{h}/m\lesssim 1.08$, the brane must be overeager in such cases.
  }
  Once some part of the brane oversteps the horizon, that part can
  never come out from the horizon by the definition of the event horizon.
  Since no equilibrium configuration as the BH
  embedding exists, the brane has no choice but to remain dynamical
  with intersecting the horizon.

\begin{figure}[tbp]
   \begin{center}
    \includegraphics[scale=0.8,clip]{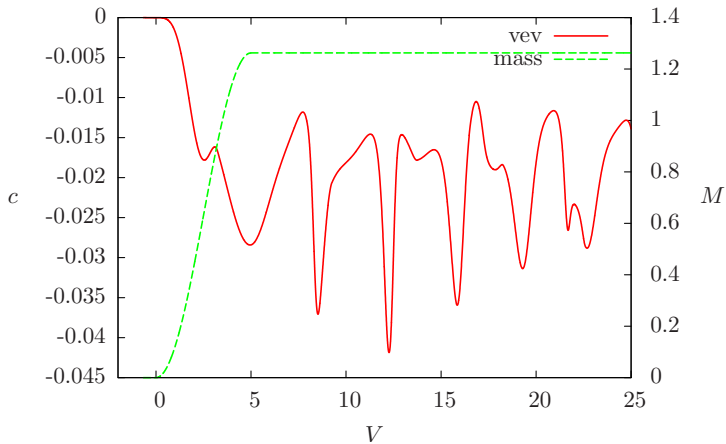}
    \caption{The quark condensate and the BH mass for duration of injection 
    $\Delta V = 5.0$ (slow injection). 
    The horizon radius at the final state is $r_\mathrm{h} = 1.06$, where the 
    equilibrium value of $c$ is $c_\mathrm{eq} = - 2.00\times 10^{-2}$.}
    \label{fig:vev_r106d5}
   \end{center}
  \end{figure}

  It is worth noting that it depends on the energy injection duration $\Delta V$ whether this overeager case is realized or not.
  As we mentioned in Sec.~\ref{subsec:subcri}, the excitations on the
  brane become larger as the energy injection becomes faster.
  Therefore, it is expected that if the energy injection is sufficiently
  slow, the brane can not be overeager. 
  In fact, we can see this expectation is true.
  Figure~\ref{fig:vev_r106d5} shows time evolution of $c$ in a case of
  slow injection.
  This reveals that the quark condensate $c$ is periodically oscillating
  around the equilibrium value, and its behavior is different from the one in
  the overeager case shown previously.
  This result implies that the brane is the Minkowski embedding with
  periodic oscillations in the sub-critical cases.
  Thus, non-adiabaticity of the energy injection plays an important role in
  the overeager effect.

Are the mesons melting in the overeager case?
As shown in Fig.~\ref{fig:vev_overeager}, 
the quark condensate $c(V)$ shows a damped oscillation around the smooth
time-evolving component, while the normal-mode oscillations 
present in the sub-critical case is absent in this case.
Also, if we focus on the snapshots of the brane configuration on constant $V$ surfaces, we find that 
the brane touches the bulk black hole for a time scale relatively longer than that of the thermalization (see Fig.~\ref{fig:snapshot_late}). 
These facts suggest that the meson is melting in the overeager case, though the temperature of the bulk black hole is lower than the critical one.
We argue the dual field theory implications of this overeager case in Sec.~\ref{sec:boundary}.


\subsection{Final fate for the overeager case}

What will be the final brane configuration of the time evolution in the overeager case?
As mentioned in the previous subsection, there is no corresponding static BH embedding.
As explained in Appendix~\ref{sec:num}, we can solve the
evolution equations only outside of the event horizon and the red-shift
prevents us from computing in the region very close to the horizon. 
For these reasons, we do not have whole numerical solutions of the
brane configuration at late time.
In this subsection, to infer a final fate of the overeager case, we extrapolate numerical solutions of the brane configuration 
using the Pad\'{e} approximation, and estimate late-time evolutions of
the brane.

\begin{figure}[tbp]
  \centering
\subfigure[Brane configuration]
{\includegraphics[scale=0.45]{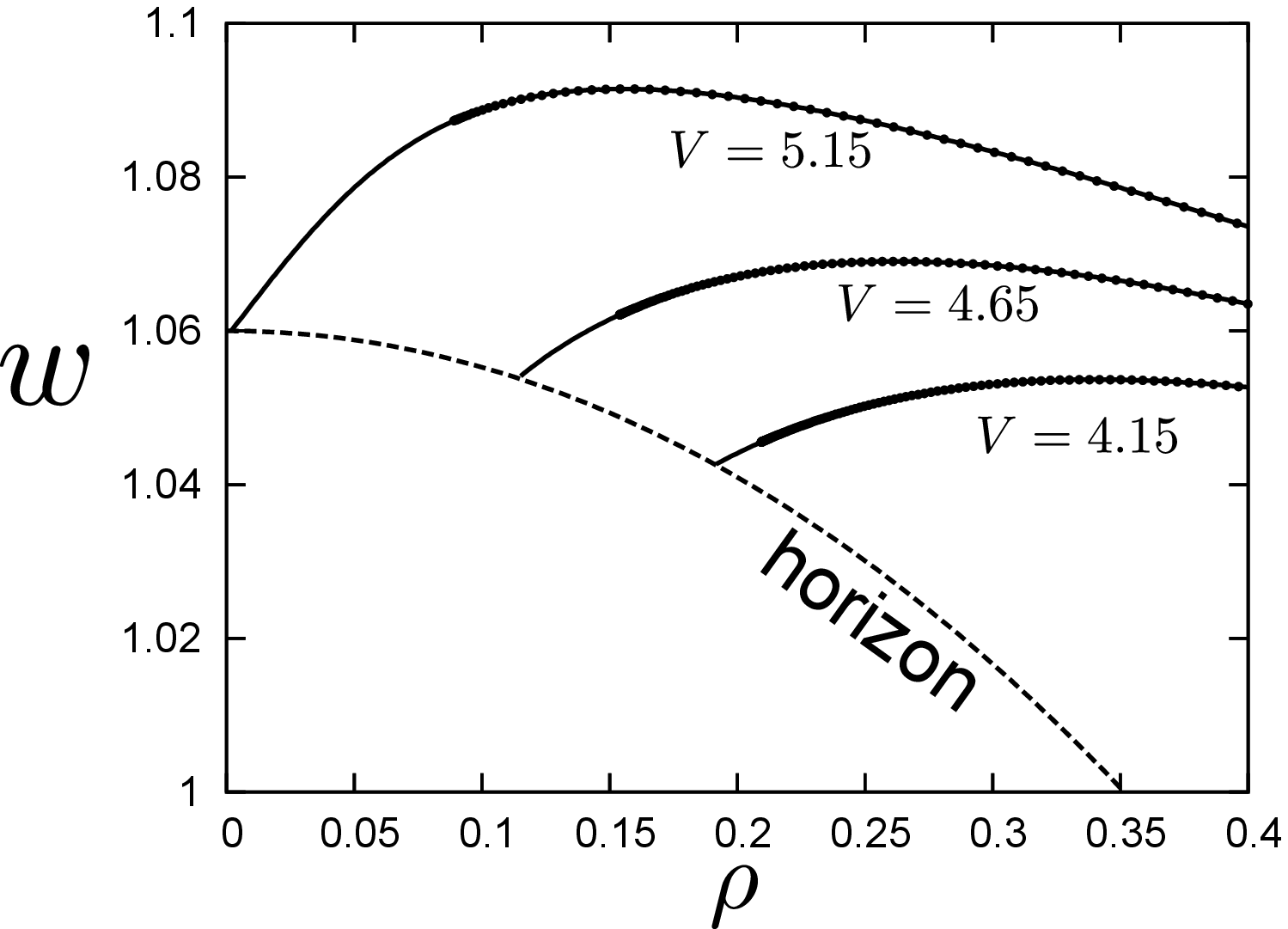}
\label{evia}
}
\subfigure[Radius of $S^3$ wrapped by the brane]
{\includegraphics[scale=0.45]{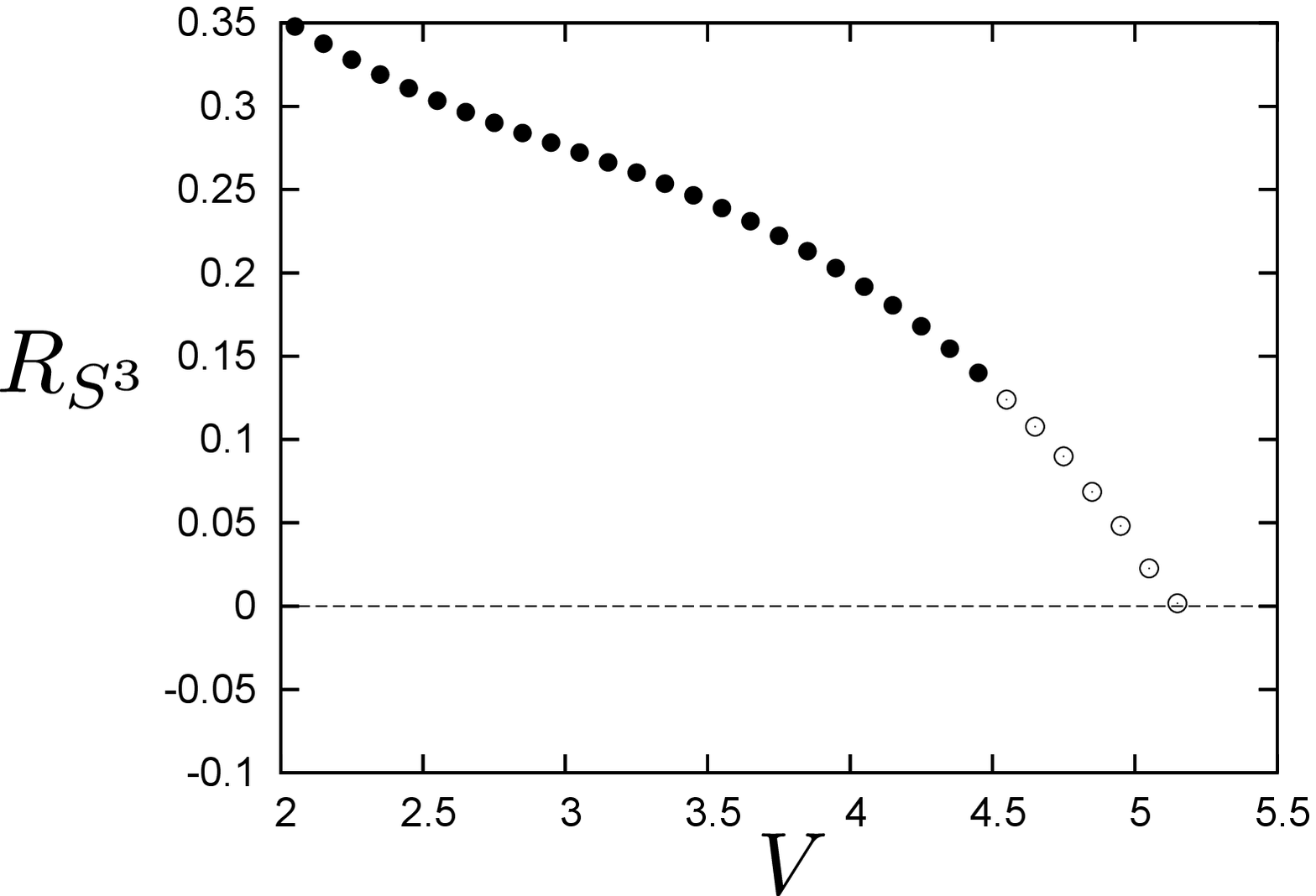}
\label{evib}
}
\caption{
(a) Snapshots of embeddings for $V=4.15,4.65,5.15$. 
Our numerical data are inter/extrapolated by solid curves.
(b) Radius of $S^3$ wrapped by the brane on
the horizon as a function of $V$.
 \label{evi}
}
\end{figure}

Let us define $e\equiv (r_\textrm{min}-r_\textrm{h})/r_\textrm{h}$
where $r_\textrm{min}$ is the minimum value of $r\equiv 1/z$ in our
numerical data at each bulk time.
This parameter $e$ measures the distance of the extrapolation.
We use numerical data satisfying $e<0.03$ in this subsection.
Figure~\ref{evia} shows the brane configuration for several late-time $V$-constant
slices:
$V=4.15,4.65,5.15$. 
We set parameters as $r_\textrm{h}=1.06$ and $\Delta V=0.5$.
Our numerical data are plotted by black dots and
their inter/extrapolation are depicted by solid curves. 
We see that, at $V=V^\textrm{crit}\simeq 5.15$, the brane position on the
horizon reaches the pole, at which the radius of the $S^3$ wrapped by the brane becomes zero.
In Fig.~\ref{evib}, we plot the radius of the $S^3$ on
the horizon, $R_{S^3}\equiv \cos \Phi|_{r=r_\textrm{h}}$, as a function
of $V$. 
In this figure, numerical data with $e<10^{-2}$ and 
$10^{-2}<e<3\times 10^{-2}$ are shown by black and white 
circles, respectively.
From the figure, we can see that the radius of $S^3$ becomes small as
time increases, and it seems to reach zero within finite time.
Since the extrapolation of the brane shape employed here may
cease to be a good approximation near the horizon and the pole when $e$
becomes large, the precise motion of the brane at late time could be different from
that described here in that case. 
However, these results with a help of the extrapolation imply
that the brane would reach the pole in the neighborhood outside the horizon
around $V \sim V^\mathrm{crit}$ and the brane embedding would be singular.\footnote{
Originally, the tip of the brane is located at the pole and the
embedding is regular there.
If the embedding is regular at $V=V^\textrm{crit}$, 
the tip of the brane must be just at the pole on the horizon.
However, the tip of the brane has fallen into the
black hole and a regular tip cannot be at the pole on the horizon.
This implies that other parts of the brane have concentrated on the pole
and the embedding would be singular.
}
Classical description of the brane would break down at such a singular
point
and a singularity outside the horizon could be seen from the boundary.

If we take into account stringy effects, 
there is a possibility that 
the brane reconnects near the pole 
around $V \sim V^\textrm{crit}$.
When
the radius of the $S^3$ wrapped by the brane becomes small,
we may expect that brane will reconnect.
Let us estimate the condition for the brane reconnection.
The radius of the $S^3$ is given by $R_{S^3}=L\cos\Phi|_{\mathrm{reconnect}} \simeq L \epsilon$
where we defined $\epsilon=\pi/2-\Phi|_{\mathrm{reconnect}}$. 
Note that we have restored the AdS curvature scale $L$, which is the radius of $S^5$ too.
The condition for the reconnection is given by $R_{S^3}\lesssim \ell_s$, and it can be rewritten as
$\epsilon\lesssim \lambda^{-1/4}$, where $\lambda$ is the 't Hooft coupling defined by $\lambda=L^4/\ell_s^4$.
This inequality indicates that,
if we take into account the stringy (finite $\lambda$) effect, emergence of the singularity may be avoided by the brane reconnection. 
In a similar manner, the reconnection may occur by taking bulk quantum effects into account via membrane nucleation.
Once the reconnection occurs,
the part of the brane connected to the boundary would be separated from the black hole and oscillate around 
the equilibrium configuration, which is the static Minkowski embedding determined by the bulk black hole mass.

\section{Discussions}
\label{sec:discussions}

\subsection{Boundary theory interpretations}
\label{sec:boundary}

We discuss boundary theory interpretations and implications of our numerical results, especially focusing on the overeager case.
In this case, the brane touches the horizon temporarily and tends to be singular in a finite time duration. If we take into account stringy or quantum effects in account, the brane may reconnect near the singular point, and go back to the Minkowski embedding with oscillations induced by the reconnection.
We consider this process in the boundary theory point of view below.

In the overeager case, the meson is melting even though the final
temperature of the gluon medium (represented by the thermal D3
background) is lower than the meson melting temperature. One possible
explanation for this phenomenon would be as follows. A rapid energy
injection drives the ambient gluon medium into a non-thermal state. In
such a state, the distribution function of the gluons will be deformed from
the thermal one governed by its temperature. If the deformed distribution function possesses sufficient amount of high-energy components, 
gluons would break bound states of quarks and melt the mesons.
It would be interesting if we could examine such a scenario from the bulk point
of view. 
For example, the spectrum of the bulk Hawking radiation corresponds to the distribution function of the gluon medium mentioned above, and can be studied using techniques of Refs.~\cite{Barcelo:2010pj,Kinoshita:2011qs}.
Such a study would provide useful information for this issue.

At the late time of the overeager case, the gluon medium approaches
thermal equilibrium and high-energy gluons decrease.
Since the final temperature is lower than the meson melting temperature,
some of the dissociated quarks and antiquarks will recombine into mesons
giving off the latent heat into the mesons and the gluon medium. The
recombination of mesons may correspond to the reconnection of the
D7-brane in the gravity side.

\subsection{Open questions and Future directions}

We have clarified some basic aspects of dynamics in the D3/D7 system 
with the aid of numerics in this paper, but we still have many open questions to be addressed.
Below, we list them along with ideas for future directions of the study.

\begin{itemize}

\item Applications to other systems:

In this paper, we have developed a numerical scheme that realizes a robust time evolution for a long duration to study the probe brane on the dynamical background spacetime.
This technique and its generalizations may have various applications to study non-linear dynamics within the context of the gauge/gravity duality.

One possibility is to apply to more general brane systems.
An example is the black hole-string system like those studied in Ref.~\cite{Gubser:2006bz}, which corresponds to the quarks moving in the thermal plasma medium. 
Our scheme will enable to consider the effect of the dynamical background and relaxation phenomena therein.
It will be interesting also
to consider brane systems that realize more realistic field theories, 
in particular those realizing confinement, such as the D4/D6 system~\cite{Kruczenski:2003uq} or the Sakai-Sugimoto model~\cite{Sakai:2004cn}.
We did not include the world-volume field strength and external fields in this paper,
and their effects deserve further investigation (see e.g.\ Ref.~\cite{Evans:2010xs} for the effect of external magnetic field).

Yet another future direction is to consider higher-cohomogeneity cases.
For example,
Ref.~\cite{Amsel:2007cw} considered a similar melting process realized by a rapidly-moving point particle modeled by the Aichelberg-Sexl solution in the bulk. 
Starting from a system whose temperature is slightly below the melting temperature and using the near-horizon approximation, 
they shown that the mesons melt in a spatially localized region whose size is larger than the scale naturally expected from the injected energy $E$ and the temperature difference $\Delta T$ from the melting temperature by a factor of ${\cal O}((E/N_c^2\Delta T)^{3/2})$.
This behavior is qualitatively similar to that of the overeager case, in the sense that the it becomes easier to melt mesons in the non-equilibrium regime.
They focused on the regime where $\Delta T/T\ll 1$, and then it will be interesting to see how this behavior changes in general cases.
Also, it may deserve to study if the phase corresponding to the overeager case, in which the mesons melt at temperature lower than the critical one, persists in this local heating case.

\item Brane motion and reconnection: 

We observed that, in the overeager case, the intersection locus of the brane and bulk horizon moves toward the pole, where a brane reconnection is expected to occur.
It would be interesting to clarify the brane dynamics after the
      reconnection assuming an explicit reconnection process 
or taking into account effects beyond the DBI action.
Dynamical processes involving a brane and a black hole have been studied in, e.g.,  Refs.~\cite{Flachi:2005hi,Flachi:2006hw,Flachi:2006ev,Flachi:2007ev} in the context of braneworld models,
and these previous results may give us implications to the physics after the brane reconnection or the quark recombination.
Also, it would be important to understand influence of the effects beyond the DBI action,
such as the curvature corrections arising from non-zero brane thickness~\cite{Carter:1994yt,Carter:1994ag,Frolov:2008ra,Czinner:2009rx,Czinner:2010hr}.

One possible way to analyze the brane dynamics in a simplified manner is to focus on the near-horizon region.
Ref.~\cite{Frolov:2006tc} studied static brane configurations in the near-horizon region.
They constructed a family of static solutions in that region and clarified its basic properties.
Also Ref.~\cite{Amsel:2007cw} focused on the near-horizon region in the D3/D7 system, and clarified the brane motion just after the local heating as explained above.
It may be interesting to study the brane dynamics in this setup, especially focusing on the regime where the reconnection occurs.

\item Brane induced geometry and fluctuations:

We may read out some extra pieces of information from the induced geometry on the D7-brane.
Fluctuations of the brane is governed by the induced geometry, since the intrinsic curvature gives rise an effective potential for those fluctuations.
By studying the relationship between the intrinsic geometry and the (quasi-)normal mode frequency of the brane fluctuations, we may achieve a better understanding on the time dependence of $c(V)$ from the bulk point of view.

In this context, it is interesting to focus on the overeager case,
where the brane turns from the BH embedding into the Minkowski embedding at some moment.
A naive expectation is that the decay rate of the quasi-normal modes becomes smaller as the transition time approaches, since it becomes zero after the transition.
The intrinsic curvature tends to singular in this process, and it may cause such a time dependence through the effective potential for the fluctuations.
It will be interesting to test this conjecture by comparing the time dependence of the brane intrinsic curvature and the brane fluctuations.
Also, this transition can be viewed as a black hole formation and subsequent evaporation in the brane induced geometry. 
It may deserve a further consideration if we can relate such an idea with the time dependence of the brane fluctuations and $c(V)$.

In Fig.~\ref{fig:ps_zh2d05}, we observed that the discrete spectrum of $c(V)$ in the Minkowski embedding shows a exponential decay at higher frequency region. We may be able to read out a characteristic scale from this exponential decay. It might be interesting to elaborate this idea further, though this kind of exponential decay in the spectrum could be simply due to basic properties of the decomposition of the initial perturbations into the normal modes at high frequency band (see e.g.\ Ref.~\cite{books/daglib/0013716} for behavior of Fourier coefficients in the high-frequency limit).

\end{itemize}

\section*{Acknowledgments}

We thank Norichika~Sago and Takashi~Hiramatsu for helpful advice about numerical calculations.
We also would like to thank Don~Marolf, Helvi Witek and Takahiro~Tanaka for valuable comments.
The work of TI is supported in part by European Union's Seventh Framework Programme under grant agreements ((FP7-REGPOT-2012-2013-1) no 316165, PIF-GA-2011-300984, the EU program ``Thales'' and ``HERAKLEITOS II'' ESF/NSRF 2007-2013 and is also co-financed by the European Union (European Social Fund, ESF) and Greek national funds through the Operational Program ``Education and Lifelong Learning'' of the National Strategic Reference Framework (NSRF) under ``Funding of proposals that have received a positive evaluation in the 3rd and 4th Call of ERC Grant Schemes''.
SK is partially supported by the JSPS Strategic Young Researcher Overseas
Visits Program for Accelerating Brain Circulation ``Deeping and Evolution of Mathematics
and Physics, Building of International Network Hub based on OCAMI''.
The work of NT is supported in part by World Premier International 
Research Center Initiative (WPI Initiative), MEXT, Japan, 
and JSPS Grant-in-Aid for Scientific Research 25$\cdot$755.

\appendix

\section{Numerical methods}
\label{sec:num}

In this appendix, we explain our numerical methods to solve the time
evolution of the D7-brane.
We need two kinds of numerical schemes depending on whether the brane intersects with the
event horizon or not.
Before and after the brane intersects with the event horizon, we use
numerical method  in subsection~\ref{before} and \ref{after}, respectively.

\subsection{Before the brane intersects with the event horizon}
\label{before}

In this section, we explain our numerical method before the brane intersects with
the event horizon.\footnote{If the brane intersects with the event horizon at 
the initial surface,
which occurs when the initial background temperature is sufficiently large,
we use the method in section~\ref{after} from the beginning.}

\subsubsection{Evolution scheme}
\label{EvSch}

\begin{figure}[tbp]
\begin{center}
\includegraphics[scale=0.3]{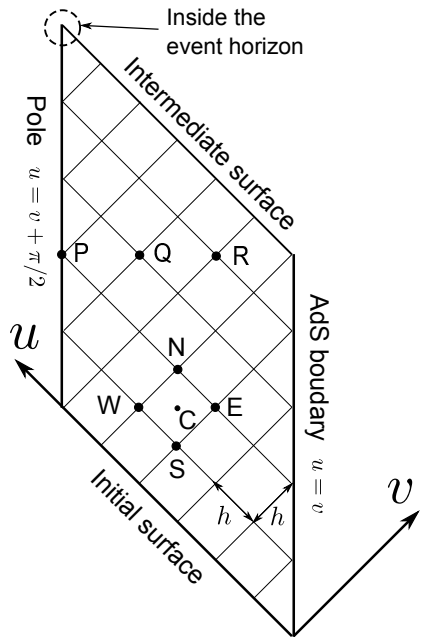}
\end{center}
\caption{Numerical domain for the world-volume of the D7-brane {\textit{before}} the brane
 intersects with the event horizon.
The AdS boundary and the pole are located at $u=v$ and $u=v+\pi/2$, respectively.
When the tip of the brane goes into the black hole, we stop our
 numerical integration and switch to the other numerical scheme shown in Sec.~\ref{after}.
}
 \label{SchMin}
\end{figure}

The time evolution of the D7-brane is described by non-linear wave
equations~\eqref{EV1}--\eqref{EV3}.
We will use the finite-difference method for solving them numerically.
It turns out that the time evolution based on these equations is numerically unstable.
To stabilize it, we eliminate $V_{,u}$ and
$V_{,v}$ in the right hand side of the equations using the
constraints~\eqref{Con1} and \eqref{Con2} as
\begin{equation}
 V_{,u}=\frac{1}{F}\biggl(
-Z_{,u}+\sqrt{Z_{,u}^2
+ FZ^2(Z\Psi)_{,u}^2
}
\biggr)\ ,
\quad
 V_{,v}=\frac{1}{F}\biggl(
-Z_{,v}+\sqrt{Z_{,v}^2
+FZ^2(Z\Psi)_{,v}^2
}
\biggr)\ ,
\end{equation}
where each sign in the front of the square root is determined by the
boundary conditions at $Z=0$, namely 
$V_{,u}=0$ and $Z_{,u}>0$, or $V_{,v}>0$ and $Z_{,v}<0$. 
As a result, 
the evolution equations can be schematically rewritten as 
\begin{equation}
 \bm{X}_{,uv}=\bm{f}(\hat{\bm{X}}_{,u},\hat{\bm{X}}_{,v},\bm{X})\ ,
\label{scheq}
\end{equation}
where $\bm{X}=(V,Z,W)$ and $\hat{\bm{X}}=(Z,W)$. 
We can realize a stable numerical time evolution if we use these equations as the evolution equations.%
\footnote{
We found the method to stabilize the numerical calculation by trial and error,
and the reason of the stabilization is unclear at this moment.
It will be useful to clarify the stabilization mechanism,
especially when we try to apply this numerical method to more generalized systems.
One way to analyze it would be to show the well-posedness of the evolution equations,
employing arguments similar to those in Ref.~\cite{Rendall08011990,Kreiss:2010ex,Babiuc:2013rra,Hilditch:2013sba}.
}
As shown in Fig.~\ref{SchMin}, we discretize $u$ and $v$ coordinates with the grid
spacing $h$. 
Now, we focus on the points N, E, W, S and C in the figure.
We can evaluate the function $\bm{X}$ and its derivatives at point C
with second-order accuracy as
\begin{equation}
\begin{split}
&\bm{X}_{,uv}|_\mathrm{C}=(\bm{X}_\mathrm{N}-\bm{X}_\mathrm{E}-\bm{X}_\mathrm{W}+\bm{X}_\mathrm{S})/h^2+\mathcal{O}(h^2)\
 ,\\
&\bm{X}_{,u}|_\mathrm{C}=(\bm{X}_\mathrm{N}-\bm{X}_\mathrm{E}+\bm{X}_\mathrm{W}-\bm{X}_\mathrm{S})/(2h)+\mathcal{O}(h^2)\
 ,\\
&\bm{X}_{,v}|_\mathrm{C}=(\bm{X}_\mathrm{N}+\bm{X}_\mathrm{E}-\bm{X}_\mathrm{W}-\bm{X}_\mathrm{S})/(2h)+\mathcal{O}(h^2)\
 ,\\
&\bm{X}|_\mathrm{C}=(\bm{X}_\mathrm{E}+\bm{X}_\mathrm{W})/2+\mathcal{O}(h^2)\ ,
\end{split}
\end{equation}
where the index of $\bm{X}$ denotes each point N, E, W or S at which
$\bm{X}$ is evaluated.
Substituting the above expressions into Eq.~\eqref{scheq}, we obtain nonlinear
equations for $\bm{X}_\mathrm{N}$. Solving the nonlinear equations by the
Newton-Raphson method or the predictor-corrector method, we can determine
unknown value of $\bm{X}_\mathrm{N}$ from $\bm{X}_\mathrm{E}$,
$\bm{X}_\mathrm{W}$ and $\bm{X}_\mathrm{S}$ which have been already given. 

\subsubsection{Boundary conditions at the AdS boundary}
\label{AdSbA}

One of the boundaries in our numerical domain is the AdS boundary, $Z=0$.
As we explained in section~\ref{sec:BCandID}, 
we fix the location of the AdS boundary at $u=v$ by using a residual
coordinate freedom throughout our calculations.
Suppose that the point C is located on $u=v$.
The boundary conditions for $Z$ and 
$\Psi$ 
leads to $Z_\mathrm{N}=0$ and
$\Psi_\mathrm{N}=1$ 
(where the quark mass $m$ is unity), shortly.
For $V$, the boundary conditions~\eqref{AdScond} are discretized as 
\begin{equation}
 V_\mathrm{N}-V_\mathrm{S}=Z_\mathrm{N}-Z_\mathrm{E}+Z_\mathrm{W}-Z_\mathrm{S}\ ,\quad
 Z_\mathrm{N}-Z_\mathrm{E}-Z_\mathrm{W}+Z_\mathrm{S}=0\ ,
\end{equation}
where the point E is a dummy grid point introduced in order to impose the boundary conditions.
From $Z_\mathrm{N}=Z_\mathrm{S}=0$ on the boundary, we obtain 
\begin{equation}
 V_\mathrm{N}=2Z_\mathrm{W}+V_\mathrm{S}\ .
\end{equation}
This equation determines the time evolution of $V$ at the AdS boundary.

\subsubsection{Boundary conditions at the pole}
The other boundary is the pole, $\Phi=\pi/2$, at which the radius of $S^3$ wrapped by
the D7-brane shrinks to zero.
We set the location of the pole at $u=v+\pi/2$.
For $Z$ and $V$, we have the Neumann boundary conditions as 
\begin{equation}
 Z_{,u}-Z_{,v}=0\ ,\quad
  V_{,u}-V_{,v}=0\ ,
\end{equation}
at $u=v+\pi/2$.
(If we introduce Cartesian coordinates $\sigma=u-v$
and $\tau=u+v$, they are rewritten as 
$Z_{,\sigma}|_{u=v+\pi/2}=V_{,\sigma}|_{u=v+\pi/2}=0$.)
Now, we focus on the points P, Q and R in Fig.~\ref{SchMin}. Since $V$ and
$Z$ satisfy the Neumann boundary conditions, we can approximate them 
by quadratic functions as $V=V_\mathrm{P}+a(u-v-\pi/2)^2$ and $Z=Z_\mathrm{P}+b(u-v-\pi/2)^2$ 
near the pole. 
We can determine $V_\mathrm{P}$, $Z_\mathrm{P}$, $a$ and $b$ from values of $V$ and $Z$ at
the points Q and R. 
As a result, we obtain
\begin{equation}
 V_\mathrm{P}=\frac{4}{3}V_\mathrm{Q}-\frac{1}{3}V_\mathrm{R}\ ,\quad
 Z_\mathrm{P}=\frac{4}{3}Z_\mathrm{Q}-\frac{1}{3}Z_\mathrm{R}\ .
\end{equation}
These expressions are essentially the backward differences of $V$ and $Z$ at the point P of second-order accuracy in the spatial direction.
For $\Psi$, we have $\Psi_\mathrm{P}=\pi(2Z_\mathrm{P})^{-1}$ 
as the boundary condition.
Note that we need to know the values of $Z$, $V$ and $\Psi$ at the points Q and R prior to imposing these boundary conditions the point P.
It is possible if we solve the evolution equations initially along
$v$-direction and then proceed to the next step along $u$-direction.  

\subsubsection{Initial data and time evolution}

We impose initial data as Eq.~\eqref{initial_data} on the initial
surface $v=0$.
Since we have located the AdS boundary and the pole at $u=v$ and
$u=v+\pi/2$, respectively, the function $\phi(u)$, which is the residual coordinate freedom, should be $\phi:[0,\pi/2] \to [0,\pi/2]$.
For simplicity, we will choose $\phi(u)=u$ here.
Now, we can solve the time evolution in the current numerical scheme
until the brane intersects with the
event horizon.   
However, if the brane has intersected with the event horizon, 
$V$ will blow up rapidly near the horizon along $v$-direction and 
our numerical calculation will be broken immediately after the intersection. 
Therefore, we should monitor values of $Z$ and $V$ at the pole for each $v$-slice.
If the pole is inside the horizon, $Z|_{u=v+\pi/2}>Z_\mathrm{EH}(V|_{u=v+\pi/2})$, 
we pause the numerical integration at a intermediate surface $v=v_\textrm{int}$
and switch to the other scheme explained in the following subsection.
Then, we will restart to integrate 
taking the numerical solution on the intermediate surface as the initial data for the new scheme.
We denote the solutions on the intermediate surface as
$V_\textrm{int}(u)=V(u,v_\textrm{int})$, 
$Z_\textrm{int}(u)=Z(u,v_\textrm{int})$ and 
$\Psi_\textrm{int}(u)=\Psi(u,v_\textrm{int})$.
On the intermediate surface, the $u$-coordinate is defined in 
$v_\textrm{int}\leq u \leq v_\textrm{int}+\pi/2$. 
The lower and upper bounds of the inequality correspond to the AdS
boundary and the pole, respectively.
Since the solutions on the intermediate surface are obtained just on grid points, 
we interpolate them by polynomials and regard them as functions in 
$u\in [v_\textrm{int},v_\textrm{int}+\pi/2]$ to construct the initial data for the new
scheme.

\subsection{Time evolution after the brane intersects with the event horizon}
\label{after}

Here, we explain the numerical method after the brane intersects with
the event horizon.
The evolution scheme and the boundary conditions at the AdS boundary are
the same as shown in Sec.~\ref{EvSch} and Sec.~\ref{AdSbA}.
At the pole, we do not need to impose any boundary conditions because it
is already inside the event horizon and is no longer contained in the current 
numerical domain.

As the initial data, we use the functions $V_\textrm{int}(u)$, $Z_\textrm{int}(u)$ and 
$\Psi_\textrm{int}(u)$,
obtained in the last subsection.
If the brane has intersected with the event horizon at an 
initial surface or if the brane will intersect shortly after the energy injection, we use
Eq.~\eqref{initial_data} as the initial data from the beginning.
As we mentioned before, $V$ at the AdS boundary blows up if the brane is close to the event horizon.
To avoid it, we will adopt a coordinate system to synchronize, at the AdS
boundary, the coordinate time on the world-volume with the asymptotic
time $V$ using the residual coordinate freedom at the initial surface.
We consider the coordinate transformation on the initial surface as
$u=\phi(\bar{u})$. Then, the $v$-coordinate is also transformed as
$v=\phi(\bar{v})$ to locate the AdS boundary at $\bar{u}=\bar{v}$.
We choose this free function $\phi$ so that $V$ and $\bar{v}$ are
synchronized up to a constant at the AdS boundary, i.e.,
$V|_{u=v}=\bar{v}+V_0$.

\begin{figure}
\begin{center}
\includegraphics[scale=0.4]{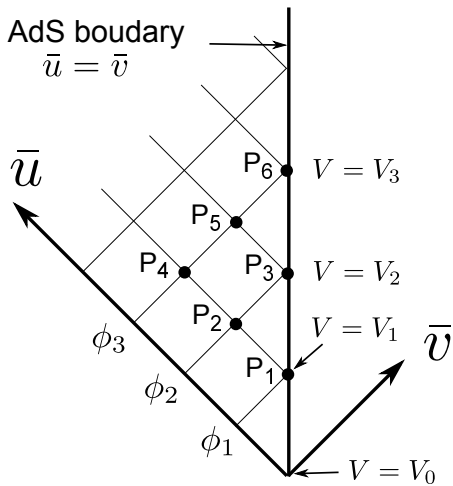}
\end{center}
\caption{Numerical domain for the world-volume of D7-brane {\textit{after}} the brane
 intersects with the event horizon.
}
 \label{SchBH}
\end{figure}

Hereafter, we explain how to determine the $\phi$ which synchronizes $V|_{u=v}$ and $\bar{v}$.
As in Fig.~\ref{SchBH}, we take the domain of $\bar{u}$-coordinate as $\bar{u}\geq 0$ and
discretize it as $\bar{u}_i=ih$ ($i=0,1,2,\ldots$). 
We denote $\phi(\bar{u}_i)=\phi_i$ and set $\phi_0=v_\textrm{fin}$. 
We also denote boundary values of $V$ at each grid point as $V_0$, $V_1$, $V_2$, $\ldots$.
We determine the numerical solution in the order of 
$\mathrm{P}_1\to \mathrm{P}_2 \to \mathrm{P}_3\to \cdots$. (Roughly speaking, we regard
$u$ as the ``time'' coordinate.)
Firstly, we solve the evolution equations at $\mathrm{P}_1$ using a trial value
$\phi_1^\textrm{trial}$.\footnote{
In our calculation, we used $\phi_1^\textrm{trial}=\phi_0+h$ and
$\phi_i^\textrm{trial}=2\phi_{i-1}-\phi_{i-2}$ for $i\geq 2$.}
Then, at the point $\mathrm{P}_1$, we obtain an asymptotic time
$V_1^\textrm{trial}$ at the AdS boundary. 
Since we would like to synchronize $V$ and $\bar{v}$ at the AdS boundary
(namely, $V_1-V_0=h$),
the appropriate choice of $\phi_1$ should be 
\begin{equation}
 \phi_1=\phi_0+(\phi_1^\textrm{trial}-\phi_0)\frac{h}{V_1^\textrm{trial}-V_0}\ .
\end{equation}
For this choice of $\phi_1$, we recalculate the evolutions and obtain $V_1=V_0+h+\mathcal{O}(h^2)$.
Next, we solve the evolution equations at $\mathrm{P}_2$ and $\mathrm{P}_3$ with a trial value
$\phi_2^\textrm{trial}$ and obtain an asymptotic time 
$V_2^\textrm{trial}$ at the AdS boundary. Since we would like to
impose $V_2-V_1=h$, we should choose $\phi_2$ as
\begin{equation}
 \phi_2=\phi_1+(\phi_2^\textrm{trial}-\phi_1)\frac{h}{V_2^\textrm{trial}-V_1}\ .
\end{equation}
Then, we recalculate them and obtain $V_2=V_1+h+\mathcal{O}(h^2)$.
By repeating this procedure, we can synchronize $V$ and $\bar{v}$ at
the AdS boundary.
Note that, in this choice of free function $\phi$, our numerical domain
does not touch the event horizon since, along a $\bar{u}$-constant surface, the
function $V$ approaches a finite value at the AdS boundary and the
surface should be outside the horizon.
(Intuitively speaking, observers at the AdS boundary cannot see the event horizon within
finite asymptotic time.)
In principle, we can solve the time evolution forever ($V|_{u=v}\to \infty$)
by the current scheme in contrast to the previous scheme in Sec.~\ref{before}.
However,
especially in the overeager case, numerical error accumulate in actual calculations 
and it becomes an obstacle for an numerical evolution over a long time duration.
In a typical set up, our numerical calculation is reliable in $V|_{u=v}-V_0 \lesssim 10$ at least,
which is sufficient for our purpose.


\section{Error analysis}

We test our numerical code by (i) checking if a static initial data remains static 
in the time evolution by our code, and (ii) showing second-order convergence of the constraints.
Our code passes these tests as shown below.

\subsection{Static initial data}

\subsubsection{Vacuum bulk}

When the bulk spacetime is the pure AdS spacetime without a black hole,
the static initial data is exactly given by Eq.~(\ref{initial_data}).
We check if our time evolution code keeps it to be static and the quark-condensate
$c(V)$ to be zero.

\begin{figure}
  \centering
  \subfigure[$|c(v)|$]
  {\includegraphics[width=5.3cm]{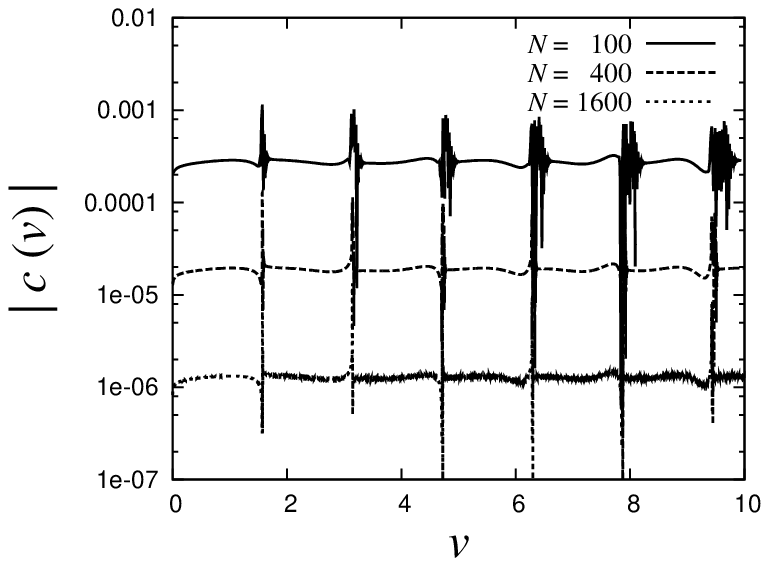}
\label{MEstatic_c}
  }
  \subfigure[Closeup of $|c(v)|$]
  {\includegraphics[width=5.3cm]{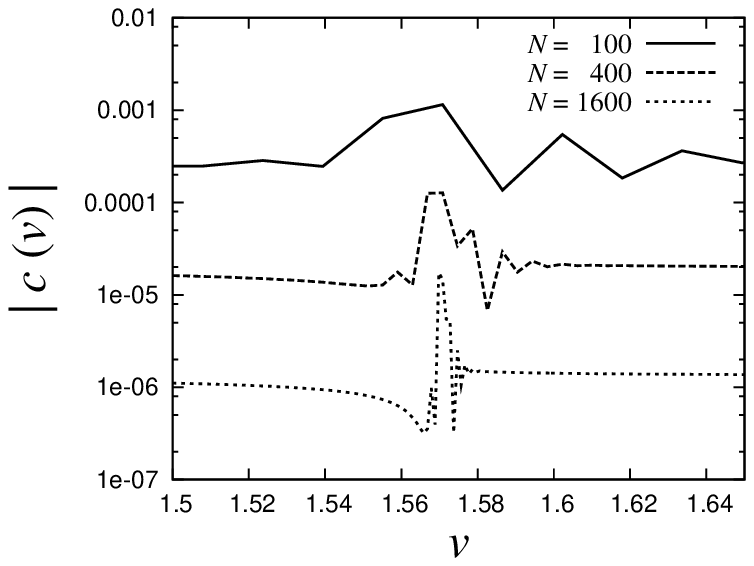}
\label{MEstatic_c_closeup}
  }
  \subfigure[$\max(|c(v)-\bar c|)$, $|\bar c|$]
  {\includegraphics[width=5.3cm]{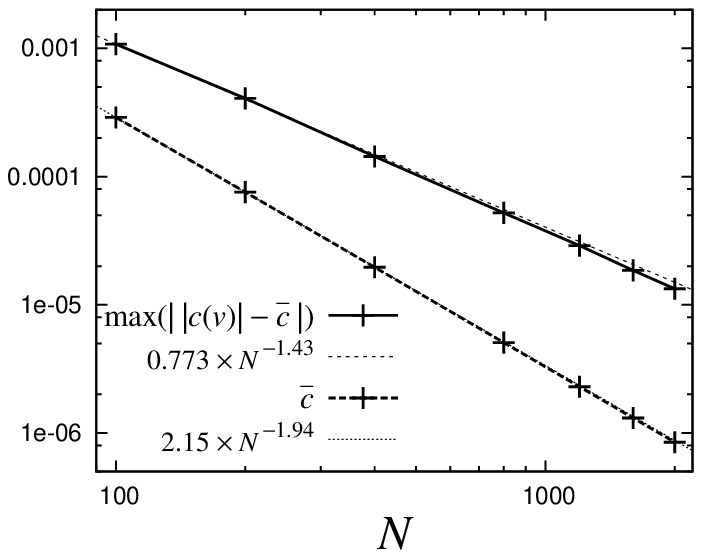}
\label{ME_blipheight}
  }
  \caption{
$|c(v)|$ for the exact static initial data~(\ref{initial_data}) in 
the pure AdS bulk spacetime.
$\bar c$ is the average value of $|c(v)|$ obtained by numerically fitting $|c(v)|$ to a constant
over $0<v<10$.
We measure the peak height from the average by $\max_{0<v<10}(\bigl||c(v)|-\bar c\bigr|)$.
}
\label{Fig:ME}
\end{figure}

We show $|c(v)|$ in this case in Fig.~\ref{MEstatic_c} for grid number 
$N=100, 400, 1600$. For each $N$, $|c(v)|$ is approximately given by an constant component with
periodic pulse noise, whose detailed structure is shown in Fig.~\ref{MEstatic_c_closeup}.
These sharp peaks are generated by time evolution near the pole, where the evolution equations become singular. 

We show the $N$ dependence of the constant component and the peak height around it in Fig.~\ref{ME_blipheight},
where we defined the average value of $|c(v)|$, $\bar c$, by numerically fitting to a constant over $0<v<10$,
and the peak height by
$\max_{0<v<10}(\bigl||c(v)|-\bar c\bigr|)$.
Fitting the numerical data, we find that both the constant component and the peak height around it converge to zero 
as we increase $N$ by power-law with powers $p=-1.94$ and $p=-1.43$, respectively.

\subsubsection{Bulk with a black hole}

We also test if static initial data remains static in the time evolution given by our code even when a bulk black hole exists.
For this purpose, we construct static initial data by integrating Eq.~(\ref{Phieqst}) numerically, and studied 
time dependence of $c(v)$ provided by that.

\begin{figure}
  \centering
  \subfigure[$c(v)$]
  {\includegraphics[width=5.3cm]{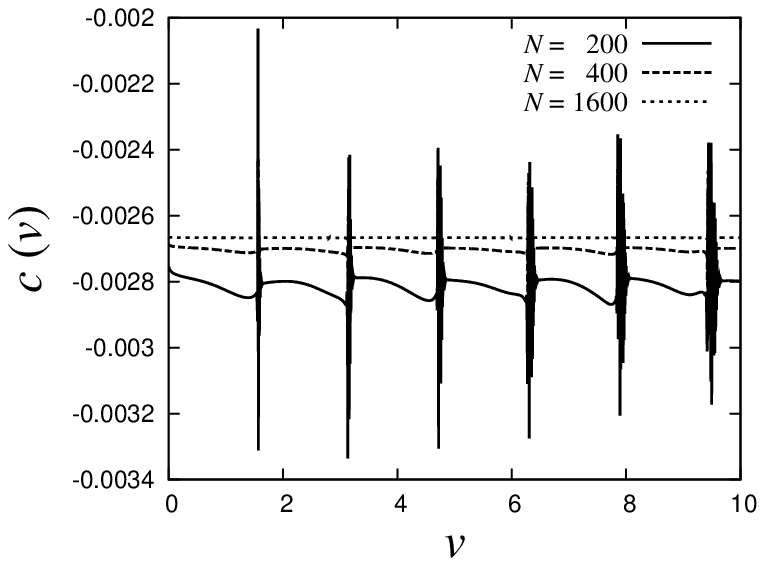}
\label{static_c}
  }
  \subfigure[Closeup of $c(v)$]
  {\includegraphics[width=5.3cm]{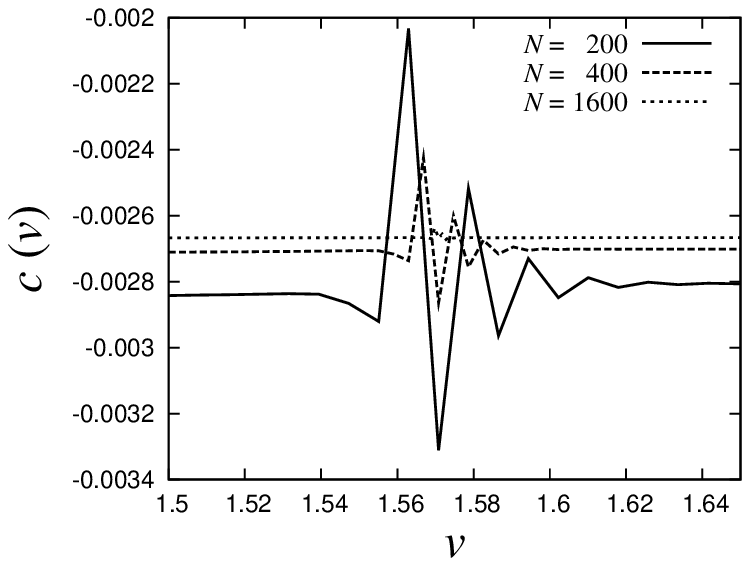}
\label{static_c_closeup}
  }
  \subfigure[$\max(|c(v)-\bar c|)$, $|\bar c(N)-c_\infty|$]
  {\includegraphics[width=5.3cm]{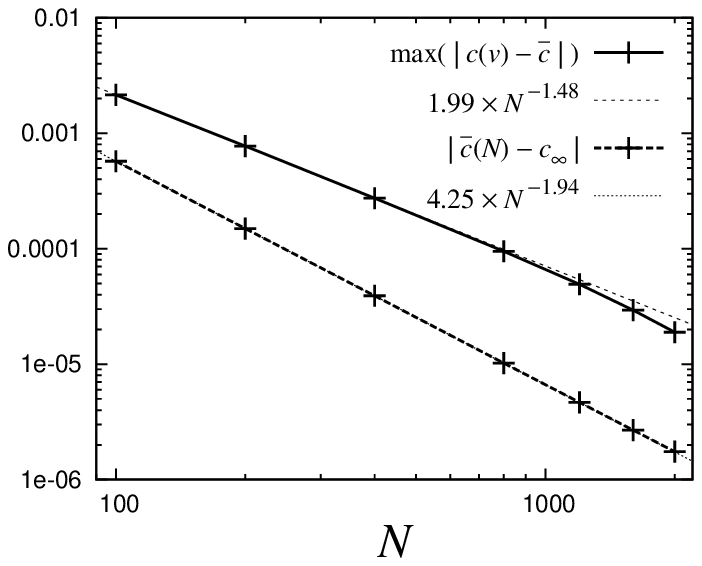}
\label{static_blipheight}
  }
  \caption{
$c(v)$ for the static Minkowski embedding in the bulk with a black hole
 with $M=0.500$ ($r_\text{h}=0.841$).
The average of $c(v)$, $\bar c$, is obtained by fitting $c(v)$ to a constant over $0<v<10$ for each $N$.
The asymptotic value of $\bar c$ is obtained by fitting $\bar c(N)=c_\infty+aN^p$, 
which gives
$c_\infty=-2.66\times 10^{-3}$ and $p=-1.94$.
}
\label{Fig:static}
\end{figure}

We show $c(v)$ for the static Minkowski embedding by Eq.~(\ref{Phieqst})
for the black hole mass $M=0.500$ (the horizon radius $r_\text{h}=0.841$) in Fig.~\ref{static_c}.
Similarly to 
the previous case,
$c(v)$ is approximately composed of constant component
and periodic pulse noise around it (see Fig.~\ref{static_c_closeup}).
We calculate the average value of the quark condensate $\bar c$ by fitting $c(v)$ to a constant over $0<v<10$
for each $N$.
This average value depends on $N$, and its asymptotic value, $c_\infty$, 
is estimated by fitting $\bar c(N)=c_\infty+aN^p$ numerically.
We show the result of the fitting in Fig.~\ref{static_blipheight}, from which 
we see that $\bar c(N)$ converges into $c_\infty = -2.66\times 10^{-3}$ by power law with $p=-1.94$.
The peak height around the average is obtained by $\max_{0<v<10}(|c(v)-\bar c|)$ for each $N$,
and it converges to zero by power law with $p=-1.48$ at least up to $N = 2000$.

\begin{figure}
  \centering
  \subfigure[$c(v)$]
  {\includegraphics[width=5.3cm]{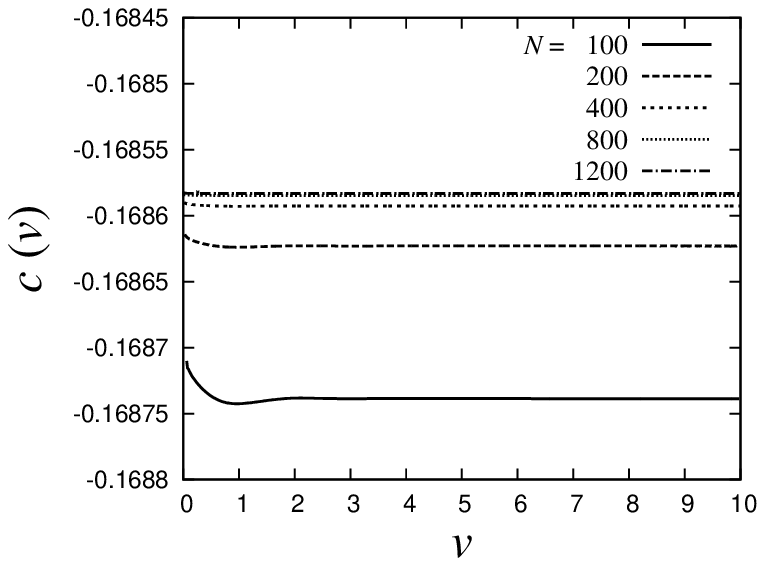}
\label{bhstatic_c}
  }
  \subfigure[$|\bar c(N)-c_\infty|$, $\max(|c(v)-\bar c|)$]
  {\includegraphics[width=5.3cm]{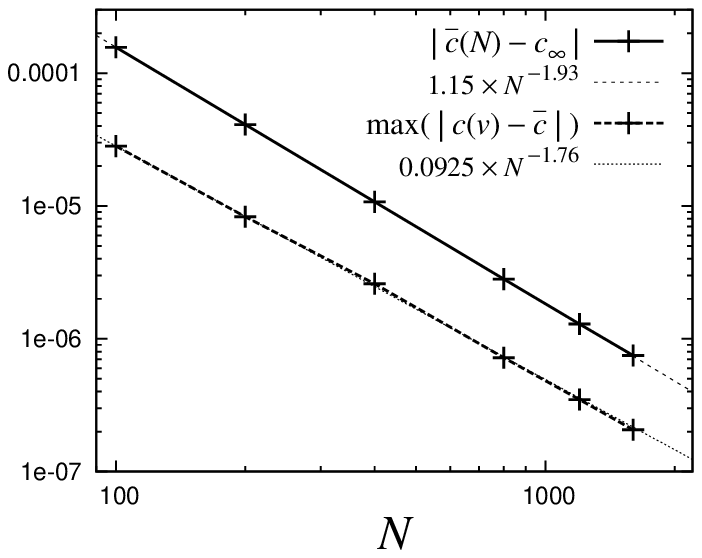}
\label{bh_c_converge2}
  }
  \caption{
$c(v)$ for the static black hole embedding in the bulk with a black hole with $r_\text{h}=1.29$.
The average $\bar c$, 
obtained by fitting $c(v)$ to a constant over $3<v<10$, 
converges to an asymptotic value
$c_\infty = -0.169$ by power law with $p=-1.93$.
The maximum deviation of $c(v)$ from the average $\bar c(N)$ converges to zero by power law
with $p=-1.76$.
}
\label{Fig:bhstatic}
\end{figure}

Next, we show $c(v)$ for the static black hole embedding for $r_\text{h}=1.29$ in Fig.~\ref{bhstatic_c}.
For each $N$, $c(v)$ converges to its final value after some oscillations around it.
Fitting shows that the average $\bar c$ for $3<v<10$ converges to an asymptotic value
$c_\infty = -0.169$ by power law with $p=-1.93$.
Also, the maximum deviation of $c(v)$ from the average $\bar c(N)$ converges to zero by power law
with $p=-1.76$.

\subsection{Constraint violation}

In this section,
we check that the constraint violation is maintained to be sufficiently small
in the dynamical calculations shown in Sec.~\ref{sec:results}.
We show that the constraint violation is sufficiently small throughout the time domain
discussed therein, and also that it shows second-order convergence when the resolution is raised.

\subsubsection{Dynamical Minkowski embedding}

Taking the numerical calculation for the results shown in Fig.~\ref{fig:vev_r106d5}
with $r_\mathrm{h} = 1.06$ and $\Delta V=5.0$ as an example, we study the resolution dependence of the constraint violation.
For this purpose, we define rescaled constraints by
\begin{align}
&\tilde C_1\equiv 
\left|
FV_{,v}^2+2Z_{,v}V_{,v}-Z^2(Z\Psi)_{,v}^2
\right|
\ ,\label{rCon1}\\
&\tilde C_2\equiv 
\left|
FV_{,u}^2+2Z_{,u}V_{,u}- Z^2(Z\Psi)_{,u}^2
\right|
\ ,
\label{rCon2}
\end{align}
by removing the overall factor $\cos^3(Z\Psi)/Z^5$ from Eqs.~(\ref{Con1}) and (\ref{Con2})
and taking their absolute values.

\begin{figure}
  \centering
  \subfigure[$\tilde C_{1,2}$ at $v=0,5,10$]
  {\includegraphics[width=5.3cm]{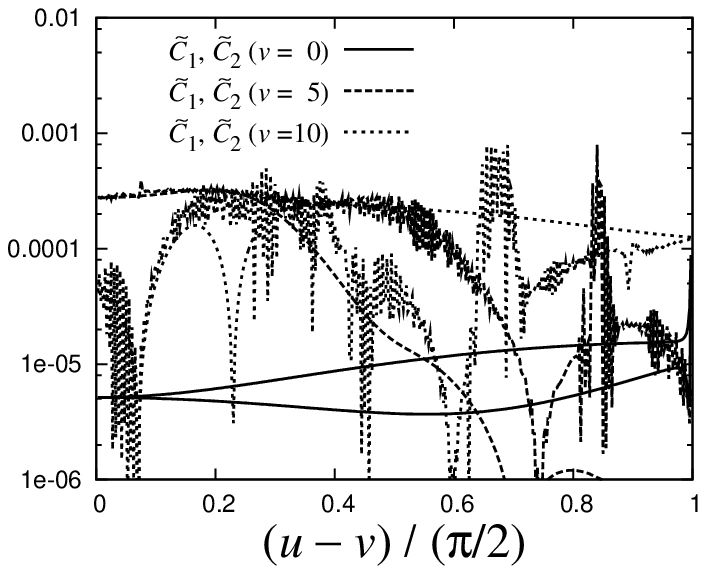}
\label{oeslow_const}
  }
  \subfigure[$\tilde C_1$ on $u-v = \pi/4$]
  {\includegraphics[width=5.3cm]{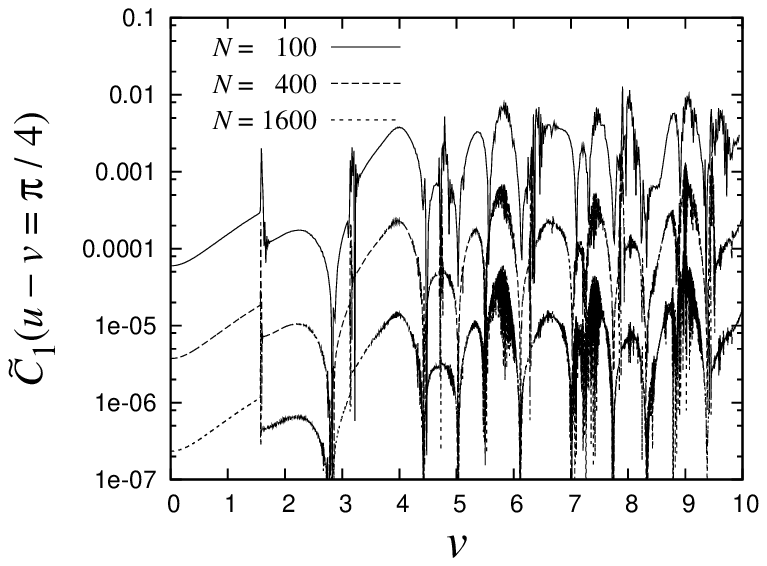}
\label{oeslow_const_line}
  }
  \subfigure[$\bar C_{1,2}(N)$ ]
  {\includegraphics[width=5.3cm]{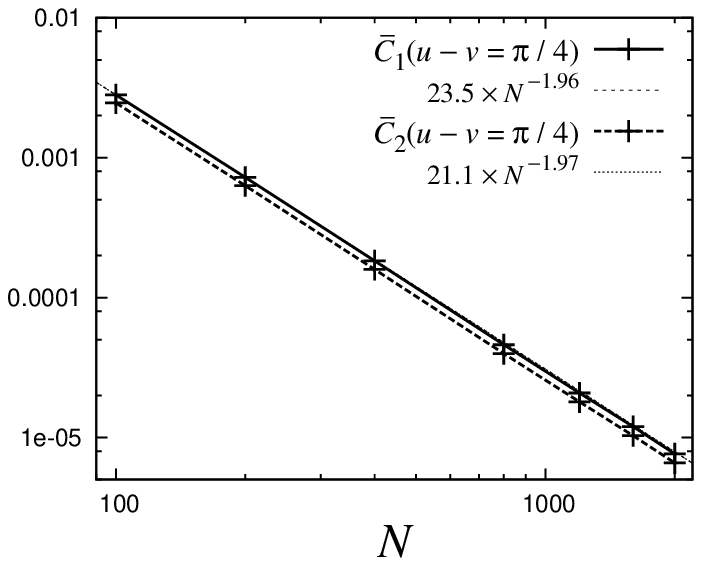}
\label{oeslow_const_converge}
  }
  \caption{
Panel (a): $\tilde C_{1,2}$ at $v=0, 5, 10$.
Panel (b): $\tilde C_1$ at $u-v = \pi/4$.
$\tilde C_2$ shows a qualitatively similar behavior.
Panel (c): Average of $\tilde C_{1,2}$ at $u-v=\pi/4$,
$\bar C_{1,2}(N)$, obtained by fitting $\tilde C_{1,2}$ to a constant over $5<V<10$.
$\bar C_{1,2}(N)$ converges to zero by power law with $p=-1.96$ and $-1.97$, respectively.
}
\label{Fig:oeslow_const}
\end{figure}

We show $\tilde C_{1,2}$ at $v=0, 5, 10$ in Fig.~\ref{oeslow_const}.
We find that $\tilde C_{1,2}<10^{-3}$ for any $u-v$ at $v=0,5,10$.
To study the time dependence more precisely, we show $\tilde C_1$
at $u-v=\pi/4$ for $0<V<10$ in Fig.~\ref{oeslow_const_line}
for $N=100, 400, 1600$. 
This plot suggests that the average of $\tilde C_1(u-v=\pi/4)$ 
roughly stays constant for $V\gtrsim 5$.
$\tilde C_2(u-v=\pi/4)$ shares a similar property.
In Fig.~\ref{oeslow_const_converge}, we show the average $\bar C_{1,2}(N)$
obtained by fitting $\tilde C_{1,2}(u-v=\pi/4)$ for $5<v<10$ to constants for various $N$.
Numerically fitting $\bar C_{1,2}(N)$, we find they converge to zero by power law with
$p=-1.96$ and $-1.97$, respectively.

\subsubsection{Dynamical black hole embedding}

Below, we show the constraint violation for the dynamical black hole embeddings discussed 
in Secs.~\ref{subsec:supcri} and \ref{subsec:oe}.

\begin{figure}
  \centering
  \subfigure[$\tilde C_{1}$ at $V=1,3,6$]
  {\includegraphics[width=5.3cm]{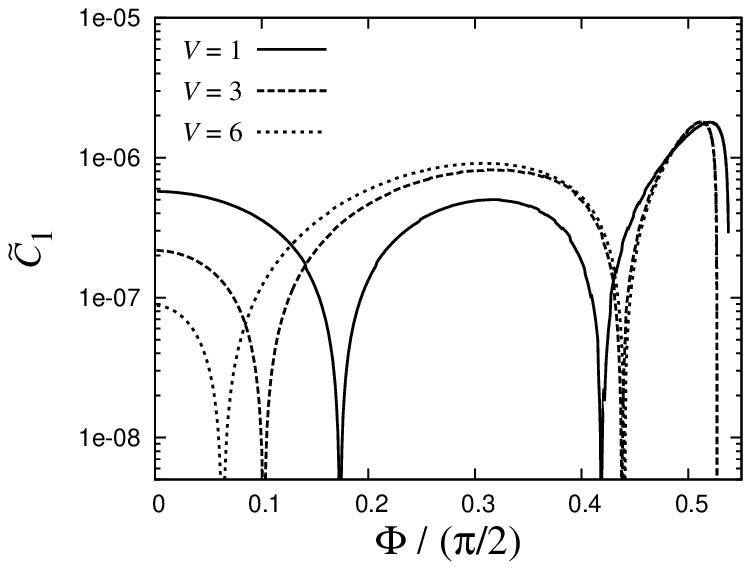}
\label{bh_const}
  }
  \subfigure[$\tilde C_1(V)$ at $Z = 10M_f^{-1/4}$]
  {\includegraphics[width=5.3cm]{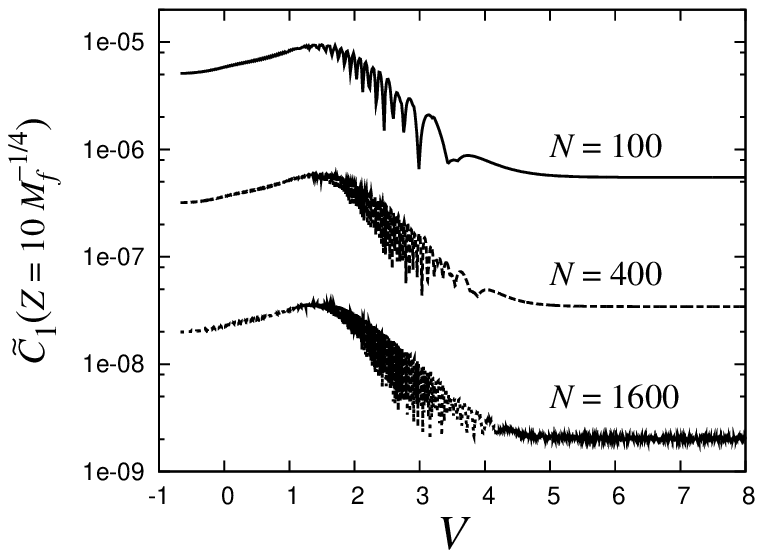}
\label{bh_const-z}
  }
  \subfigure[$\bar C_{1,2}(N)$ at $Z = 10M_f^{-1/4}$]
  {\includegraphics[width=5.3cm]{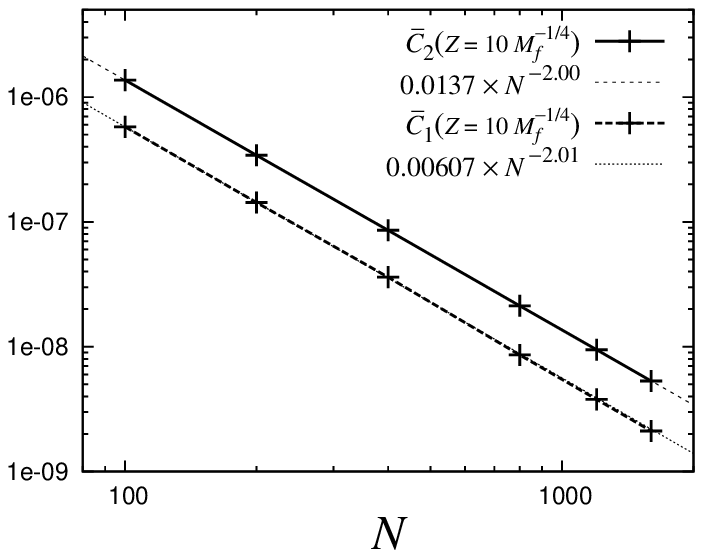}
\label{bh_const-z_converge}
  }
  \caption{
Constraint violation for 
the super-critical case
with $r_\text{h}=1.25$ and $\Delta V=1.0$.
Panel (a): $\tilde C_1$ at $V=1,3,6$. 
$\Phi$ on the horizon converges to $0.53\times\pi/2$ at late time, 
and our computational domain is limited to $0<\Phi<\Phi_\text{horizon}$.
Panel (b): $\tilde C_1$ at $Z = 10M_f^{-1/4}$.
$\tilde C_2$ shows a qualitatively similar behavior.
Panel (c): Average of $\tilde C_{1,2}$ for $4<V<8$, $\bar C_{1,2}(N)$.
They converge to zero by power law with $p=-2.01$ and $-2.00$, respectively.
}
\label{Fig:bh_const}
\end{figure}

Figure~\ref{Fig:bh_const} shows the constraint violation for 
the super-critical case 
with $r_\text{h}=1.25$ and $\Delta V=1.0$ dealt in Sec.~\ref{subsec:supcri}.
$\tilde C_1$ at $V=1,3,6$ and that at $Z=10 M_f^{-1/4}$ are plotted in Figs.~\ref{bh_const} and \ref{bh_const-z}, respectively.
Plots for $\tilde C_2$ become qualitatively similar to them.
The average value of $\tilde C_{1,2}$ at $Z=10 M_f^{-1/4}$ over $4<V<8$, $\bar C_{1,2}$, converges to zero
by power law with $p=-2.00$ and $-2.01$, respectively.

\begin{figure}
  \centering
  \subfigure[$\tilde C_{1}$ at $V=0.5, 5.5, 6.0, 7.0$]
  {\includegraphics[width=5.3cm]{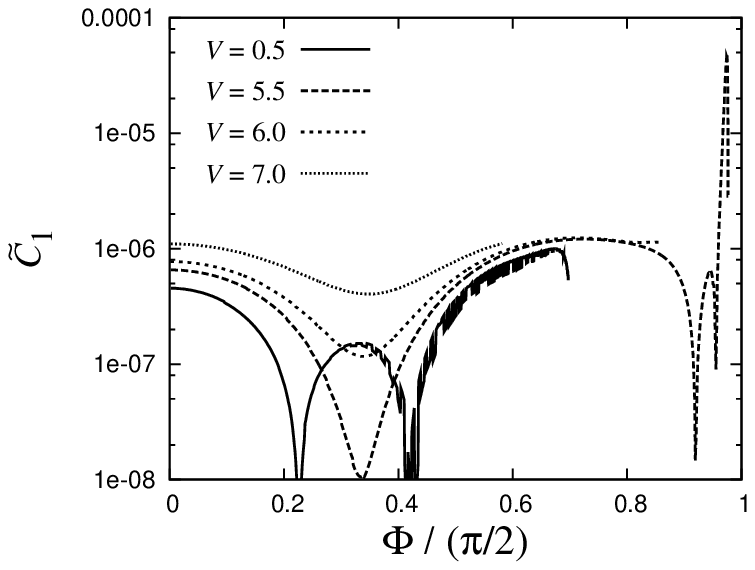}
\label{oe_const}
  }
  \subfigure[$\tilde C_1(V)$ at $Z = 10M_f^{-1/4}$]
  {\includegraphics[width=5.3cm]{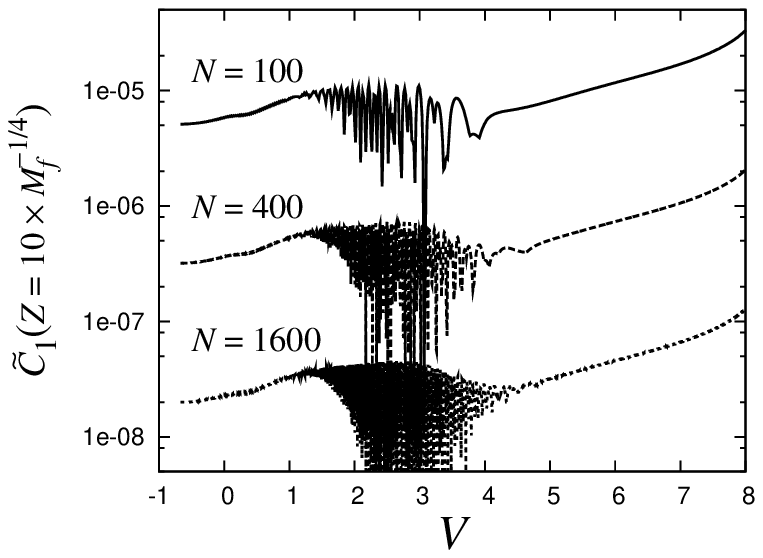}
\label{oe_const-z}
  }
  \subfigure[$\tilde C_{1,2}(N)$ at $Z = 10M_f^{-1/4}$ and $V=8$]
  {\includegraphics[width=5.3cm]{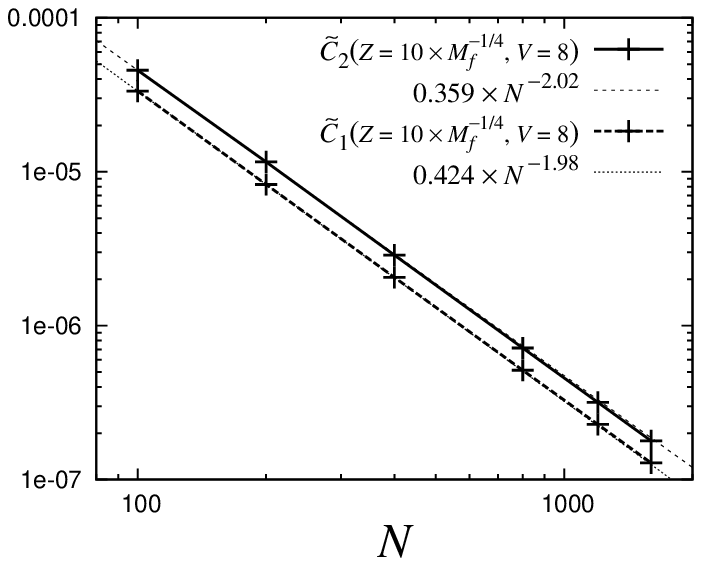}
\label{oe_const-z_converge}
  }
  \caption{
Constraint violation for the overeager case with $r_\text{h}=1.06$ and $\Delta V=0.5$.
Panel~(a): $\tilde C_1$ at $V=0.5, 5.5, 6.0, 7.0$. 
$\Phi$ on the horizon approaches $\pi/2$ at $V\simeq 5.5$, and $\tilde C_1$ 
grows to ${\cal O}(10^{-4})$ at $\Phi\simeq 0.97\times \pi/2$ for $N=100$.
Panel (b): $\tilde C_1$ at $Z = 10M_f^{-1/4}$.
$\tilde C_2$ shows a qualitatively similar behavior.
Panel (c): $\tilde C_{1,2}$ at $Z = 10M_f^{-1/4}$ and $V=8$.
They converge to zero by power law with $p=-1.98$ and $-2.02$, respectively.
}
\label{Fig:oe_const}
\end{figure}

Figure~\ref{Fig:oe_const} shows the constraint violation for the overeager case 
with $r_\text{h}=1.06$ and $\Delta V=0.5$ dealt in Sec.~\ref{subsec:oe}.
Figure~\ref{bh_const} shows $\tilde C_1$ at $V=0.5, 5.5, 6.0, 7.0$, where 
$\Phi$ on the horizon approaches $\pi/2$ at $V\simeq 5.5$
We can see that $\tilde C_1$ 
grows to ${\cal O}(10^{-4})$ at $\Phi\simeq 0.97\times \pi/2$ for $N=100$.
Figure~\ref{bh_const-z} shows $\tilde C_1$ at $Z=10 M_f^{-1/4}$, 
in which we can see that $\tilde C_1$ grows exponentially in time.
Plots for $\tilde C_2$ are qualitatively similar with them.
To see the resolution dependence, we plot $\tilde C_{1,2}$ at $Z=10 M_f^{-1/4}$ and $V=8$
in Fig.~\ref{oe_const-z_converge}.
Fitting shows that they converge to zero by power law with $p=-1.98$ and $-2.02$, respectively.
It implies that the constraint violation up to $V\leq 8$ shows second-order convergence 
to zero for higher resolution.

\providecommand{\href}[2]{#2}\begingroup\raggedright\endgroup

\end{document}